\documentclass[fleqn,usenatbib]{mnras}

\usepackage{newtxtext,newtxmath}

\usepackage[T1]{fontenc}

\DeclareRobustCommand{\VAN}[3]{#2}
\let\VANthebibliography\thebibliography
\def\thebibliography{\DeclareRobustCommand{\VAN}[3]{##3}\VANthebibliography}

\usepackage{graphicx}	
\usepackage{amsmath}	

\usepackage{subcaption}
\usepackage{xcolor}
\newcommand{\rcd}{r_{\rm cd}}
\newcommand{\rid}{r_{\rm id}}
\newcommand{\rsp}{r_{\rm sp}}
\newcommand{\rvir}{r_{\rm vir}}
\newcommand{\rta}{r_{\rm ta}}

\graphicspath{{Figures/}}

\title[Measurement of characteristic depletion radius]{First measurement of the characteristic depletion radius of dark matter haloes from weak lensing}

\author[M. Fong et al.]{
Matthew Fong,$^{1,2,3}$\thanks{matthew.fong@sjtu.edu.cn}
Jiaxin Han,$^{1,2,3}$\thanks{jiaxin.han@sjtu.edu.cn}
Jun Zhang,$^{1,2,3}$ \thanks{betajzhang@sjtu.edu.cn}
Xiaohu Yang,$^{1,2,3,4}$
Hongyu Gao,$^{1,2,3}$
\newauthor
Jiaqi Wang,$^{1,2,3}$
Hekun Li,$^{1,2,3}$
Antonios Katsianis,$^{1,2,3,4}$
Pedro Alonso$^{1,2,3}$
\\
$^{1}$ Department of Astronomy, School of Physics and Astronomy, Shanghai Jiao Tong University, Shanghai 200240, China\\
$^{2}$ Key Laboratory for Particle Astrophysics and Cosmology (MOE), Shanghai 200240, China\\
$^{3}$ Shanghai Key Laboratory for Particle Physics and Cosmology, Shanghai 200240, China\\
$^{4}$ Tsung-Dao Lee Institute, Shanghai Jiao Tong University, Shanghai 200240, China\\
}

\date{Accepted XXX. Received YYY; in original form ZZZ}

\pubyear{2021}

\begin{document}
\label{firstpage}
\pagerange{\pageref{firstpage}--\pageref{lastpage}}
\maketitle

\begin{abstract}
We use weak lensing observations to make the first measurement of the characteristic depletion radius, one of the three radii that characterize the region where matter is being depleted by growing haloes. The lenses are taken from the halo catalog produced by the extended halo-based group/cluster finder applied to DESI Legacy Imaging Surveys DR9, while the sources are extracted from the DECaLS DR8 imaging data with the \textsc{Fourier\_Quad} pipeline. We study halo masses $12 < \log ( M_{\rm grp} ~[{\rm M_{\odot}}/h] ) \leq 15.3$ within redshifts $0.2 \leq z \leq 0.3$. 
The virial and splashback radii are also measured and used to test the original findings on the depletion region. When binning haloes by mass, we find consistency between most of our measurements and predictions from the \textsc{CosmicGrowth} simulation, with exceptions to the lowest mass bins. The characteristic depletion radius is found to be roughly $2.5$ times the virial radius and $1.7 - 3$ times the splashback radius, in line with an approximately universal outer density profile, and the average enclosed density within the characteristic depletion radius is found to be roughly $29$ times the mean matter density of the Universe in our sample. 
When binning haloes by both mass and a proxy for halo concentration, we do not detect a significant variation of the depletion radius with concentration, on which the simulation prediction is also sensitive to the choice of concentration proxy. We also confirm that the measured splashback radius varies with concentration differently from simulation predictions.
\end{abstract}




\section{Introduction}

Dark matter perturbations of the primordial Universe are the seeds for the Large-Scale Structure (LSS) we observe today. Galaxies form within the growing dark matter clumps called haloes. The structure and evolution of the Universe can be modelled using dark matter haloes as the fundamental building blocks, with the halo boundary being one of the main descriptors. 

However, a single halo can be asymmetric and embedded within a complex environment such that there is no clear boundary. Attempts to describe a halo edge in theory have been made through variations of the virial radius, derived from spherical collapse~\citep{Gunn1972}.
Unfortunately, in practice these do not clearly separate the virialized region of haloes from the inflow region~\citep[e.g.][]{Cuesta2008,Zemp14} and suffer from pseudo-evolution~\citep{Diemer13}. Recently a natural and physically meaningful halo radius has been introduced, relating halo accretion to the halo boundary, called the splashback radius~\citep{Diemer2014,Adhikari2014,More2015,Shi2016}. It has proven to be a powerful tool that can potentially probe halo properties that have previously been hidden and has been measured in observations~\citep{More2016,Snaith2017,Umetsu2017,Baxter2017,Mansfield2017,Okumura2017,Fong2018,Chang2018,Okumura2018,Contigiani2019,Xhakaj2020,Sugiura2020,Aung2020,Murata2020,Deason2020,Deason2020_galaxyEdge,Zhang2021,Oneil2021}.

Another recently discovered physically motivated boundary to describe haloes is the depletion region~\citep{Fong2021}. This is where matter is experiencing a competition between the gravitational pull from the host halo and that from neighbouring haloes, in addition to the expansion of the Universe causing mass to outflow (Hubble flow). The combined effect is seen as a draining of matter on the outskirts of growing haloes. This region of depletion can be characterized by three physically meaningful radii, the inner and characteristic depletion radii as well as the turnaround radius, and provides a new way to study the evolution of haloes. The depletion radii have been shown to carry additional information than what the virial and splashback radii already do, and thus can give new insights into halo evolution and properties. 

\begin{figure} 
	\includegraphics[width=\columnwidth]{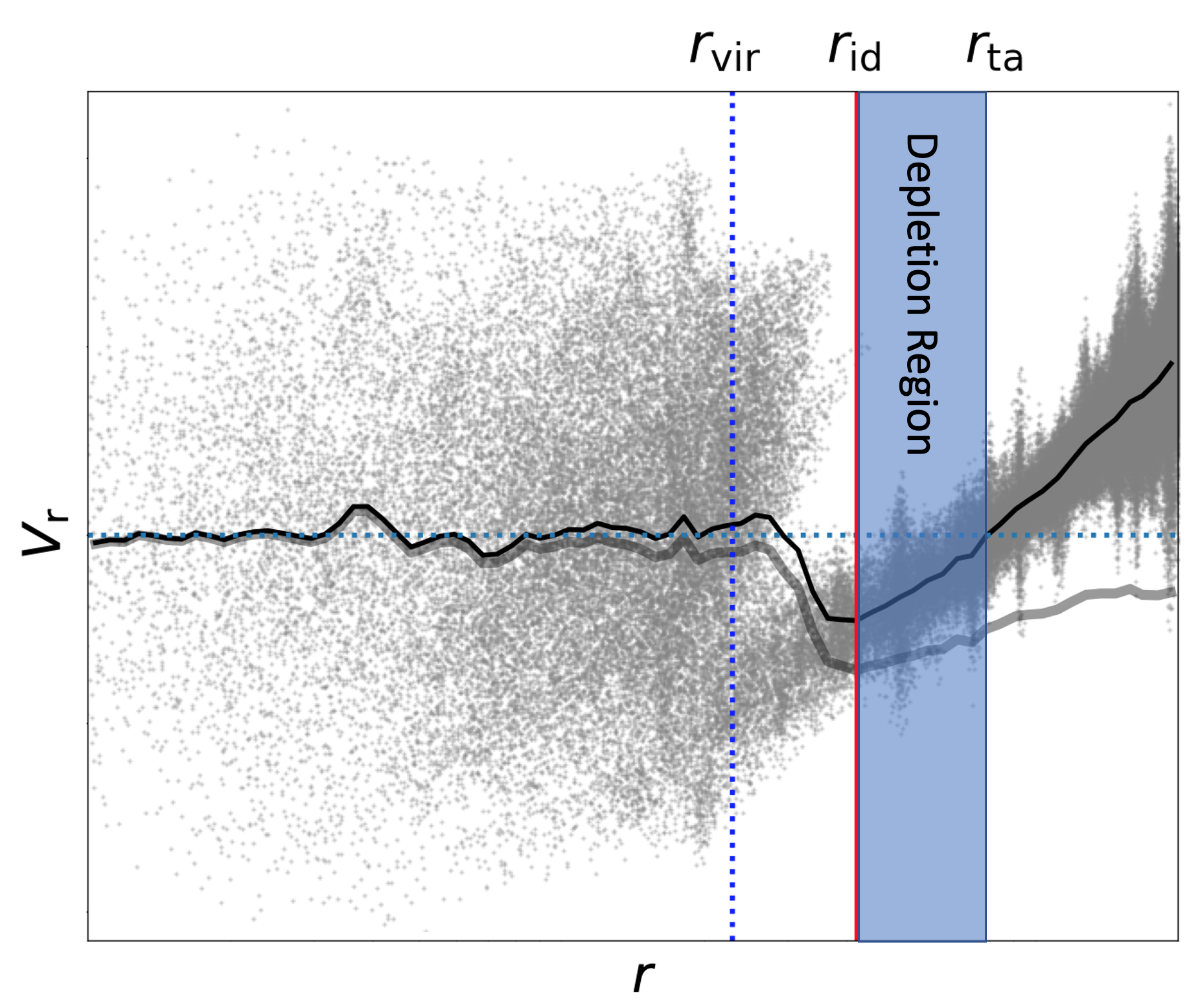}
	\caption{
	Visualization of the depletion region in the context of radial velocity, $V_{\rm r}$. The grey points are the distribution of particles around a halo of mass $M_{\rm vir} \approx 1.35 \times 10^{15} ~{\rm M_{\odot}}/h$, with only a random $1$ percent of particles shown. The solid black and grey lines are the mean radial and peculiar velocities. The depletion region is bound by the inner depletion (maximum infall or inflow location) and turnaround radii, $\rid$ and $\rta$, respectfully. Typically the splashback radius is roughly around or greater than the virial radius, while the characteristic depletion radius is within the depletion region.
	}
	\label{fig:cartoon_depletionRegion}
\end{figure}

We summarize the explicit definitions of the various aforementioned halo radii below.
\begin{itemize}
    \item The virial radius, $\rvir$, is the expected radius where the matter within is virialized according to the spherical collapse model. Normally this is defined through the expected virialization density, which we take from the prediction of~\citep{Bryan1998} assuming a tophat initial density in a spherical collapse model. Variations of the virial radius, such as $r_{\rm 200m}$ or $r_{\rm 200c}$, have been used to describe the halo mass and is traditionally used in constraining cosmology.
    
    \item The splashback radius, $\rsp$, is traditionally located where the density profile reaches its steepest slope. The steepening in the slope has been attributed to the build-up of particles during their first orbital apogees, where the particles have low radial velocities~\citep{Fillmore84,Bertschinger85}. This radius has gathered significant interest since its discovery as it has been shown to probe the mass accretion rate of haloes~\citep{Diemer2014, Adhikari2014, More2015, Shi2016}. Note the splashback radii of individual halo particles can span a wide range, and hence there is a large freedom in defining the overall splashback radius of a halo from particle dynamics~\citep{Mansfield2017,Diemer2017A, Diemer2017B,Diemer2021}. In this work, we will use the steepest logarithmic density slope location as the splashback radius, unless explicitly specified otherwise. 
    
    \item The inner depletion radius, $\rid$, is located where matter is typically flowing most quickly towards the halo centers. In simulations this also corresponds to the maximum inflow location, the minimum of the mean radial velocity or mass flow rate profiles.\footnote{When haloes are binned by mass these locations are found to be very similar~\citep{Fong2021}.} According to continuity, the maximum inflow radius marks the transition between an inner growing halo and the depleting environment, or the inner edge of the active depletion region. Therefore, this location separates the growing halo from its depleting environment and matches very well with the optimized exclusion radius used in halo modelling~\citep{Garcia2020,Fong2021}. In the splashback context, this can be interpreted as a radius that encloses a highly complete sample of splashback particles, due to the contribution of the backsplashing particles that counter the infall stream in the radial velocities (see Figure~\ref{fig:cartoon_depletionRegion}).
    
    \item The characteristic depletion radius, $\rcd$, is located at the halo bias minimum and represents the location where the clustering around haloes become the weakest relative to the underlying overdensity of the Universe. 
    The existence of the bias minimum can be understood as the result of the mass depletion around haloes, creating a relative draining of matter starting from the inner depletion radius out to the turnaround radius, and is located between the two (see Figure~\ref{fig:cartoon_depletionRegion}). The characteristic depletion radius can be used with the virial, splashback, and turnaround radii to give new insights into the outer density profile and the accretion phase of haloes~\citep{Fong2021}. 
    
    \item The turnaround radius, $\rta$, is located where the radial velocity of particles around haloes have typically just been overcome by the Hubble flow (see Figure~\ref{fig:cartoon_depletionRegion}). The combination of gravity and dark energy impact on particle velocities, and thus the turnaround radius can potentially be used as a cosmological probe~\citep{Taruya2000,Mo2010,Pavlidou2014,Falco2014, Faraoni2015,Tanoglidis2015,Tanoglidis2016, Lee2017_rta, Korkidis2019}. In the context of halo depletion, this can be used to characterize the outer edge of the depletion region. 
\end{itemize}

We plot an example of the depletion region in Figure~\ref{fig:cartoon_depletionRegion}, bound by the inner depletion and turnaround radii. The depletion region can be characterized by three radii: the inner depletion radius ($\rid$), the characteristic depletion radius ($\rcd$), and the turnaround radius ($\rta$). 
Though not shown here for clarity, the splashback radius is typically around or greater than the virial radius while the characteristic depletion radius is within the depletion region. 
We \textit{typically} find $\rvir \lesssim \rsp < \rid \lesssim \rcd \lesssim \rta$, though this relationship depends on the halo properties being studied.

The characteristic depletion radius is rich in information and has intriguing properties~\citep{Fong2021}.
When binning haloes at redshift $z=0$ by virial mass we find that $\rcd = 2.5 \, \rvir$, and $\rho(\leq \rcd) = 40 \, \rho_{\rm m}(z=0)$, where $\rho(\leq \rcd)$ is the density enclosed within $\rcd$ and $\rho_{\rm m}(z=0)$ is the matter density of the Universe today ($z=0$). These results suggest that when haloes are binned by mass there are simple ways to predict the characteristic depletion radius and that the enclosed density is a constant with mass.
The characteristic depletion radius is roughly $2 - 3$ times the splashback radius.
These combined give clues to a mostly universal density profile, where deviations from universality is constrained and may be due to accretion history.
Using the turnaround radius with the depletion radii we can also gain insights into the accretion phase of haloes on depletion scales.
When binning by two halo properties $\rcd$ is found to have different halo property information than what the splashback radius already has, and both hold more information beyond the traditional virial radii definitions. 
We hope that combining halo boundaries can be used in understanding halo features that may have been previously hidden. 

The depletion radius is located in between the halo scale and the linear scale of matter growth. This intermediate scale has already been studied in many previous works both theoretically and observationally~\citep[e.g.,][]{Cooray2002,Tasitsiomi2004,Weinberg2004,Sheldon2004,Mandelbaum2005,Prada2006,Hayashi08,Fong2018,Garcia2020,Fong2021,Diemer2021}.
However, the characterization of the depletion region is new and the potential uses have yet to be fully realized. Thus observations of these radii have not been explored until recently. The first measurement of the inner depletion radius has been found using the radial velocities of satellites around the Milky Way~\citep{Li2021}.
In this work we make the first measurement of the characteristic depletion radius using weak gravitational lensing. We also measure the splashback and virial radii and compare our results with simulation predictions, obtaining multiscale characterizations of the halo sizes.

This paper is organized as follows. In Section~\ref{sec:observations} we discuss the weak lensing measurement pipeline as well as the source and lens catalogs used in this work. In Section~\ref{sec:simulation} we introduce the simulation we use in predicting the characteristic depletion radius measurements. In Section~\ref{sec:biasAndWeakLensing} we discuss the method to measure $\rcd$ through the bias profile, that can be measured in weak lensing. In Section~\ref{sec:results} we examine our findings when haloes are binned by a single halo parameter, mass, and two halo parameters, mass and concentration. In Section~\ref{sec:systematics} we include a few potential systematics in our measurement.
Summary and conclusions are in Section~\ref{sec:conclusions}.

Note that in this paper $\log$ is in base $10$, all radii are in physical scales with units ${\rm Mpc}/h$, the masses plotted are the weak lensing inferred masses following the virial definition of~\citet{Bryan1998}, masses are in units of ${\rm M_{\odot}}/h$, and the results are all given in the cosmology discussed in Section~\ref{sec:simulation}, unless stated otherwise. We build our bias profile weak lensing module inspired by the \textsc{colossus} package~\citep{Diemer2018} and use the \textsc{colossus} package for most cosmological calculations throughout this work.

\section{Observations}
\label{sec:observations} 
In this section we briefly introduce the weak lensing pipeline, source catalog, and lens catalog used in this work. The publicly available group/cluster, shear, and mock DESI galaxy and group catalogs can be found in \url{https://gax.sjtu.edu.cn/data/DESI.html}.

\subsection{Weak lensing measurement and source catalog}
\label{sec:weakLensingMeasurements}

Our weak lensing measurements are based on the shear catalog constructed from the Dark Energy Camera Legacy Survey (DECaLS) DR8 imaging data~\citep{Dey2019} using the \textsc{Fourier\_Quad} (FQ) shear measurement pipeline~\citep{Zhang2008,Mandelbaum2015}. 
The FQ method defines the shear estimators with the multipole moments of the galaxy power in Fourier space. In this method, the effects due to the Point Spread Function, the background noise, and the source Poisson noise are all corrected in model-independent ways, and tested under general observing conditions to a very low Signal-to-Noise Ratio (SNR) of the source image (SNR $< 10$)~\citep{Zhang_2015}. More recently, it has been shown using both the CFHTLenS~\citep{Heymans_2012,Erben_2013} and DECaLS data that the FQ shear estimators of the galaxies can faithfully recover the small ($\sim 10^{-3}$) field distortion signals caused by the optics of the telescope~\citep{Zhang2019,Wang_2021}. In doing so, we caution that inappropriate selection of sources due to either the selection function itself or the presence of the geometric boundaries (CCD edges, bad pixels, etc.) could lead to systematic errors in shear recovery. These errors have all been specifically discussed and explicitly removed~\citep{Li_2021,Wang_2021}. The FQ image processing pipeline for the DECaLS data is only slightly different from that of~\cite{Zhang2019}. Some details are reported in a separate work (Zhang et al., in preparation).

Our DECaLS shear catalog covers about 9000 square degrees in g/r/z bands. As a result of the field distortion test, we find that the g-band data is not as good as the other two bands in terms of image quality. Therefore, in this work, we only use the r and z band data. In FQ, images of the same galaxy on different exposures would be treated as independent images, and produce independent shear estimators. On average, we achieve more than 10 source images per square arcmin, with a median redshift at around 0.55. The photometric redshift source catalog is from~\cite{Zou2019}, and in the FQ pipeline, we only consider the background source redshifts larger than $0.2$ beyond the foreground redshift.

To measure the excess surface density around the foreground lens, we adopt the PDF-symmetrization (PDF\_SYM) method developed in~\cite{Zhang2017}. The PDF\_SYM method is developed for the purpose of fully utilizing the information in the ensemble of the shear estimators, so that the resulting statistical uncertainties becomes minimum, i.e., the Cramer-Rao Bound. \cite{Zhang2017} find that this can indeed be achieved with the FQ shear estimators under general observing conditions. The main procedure is to find the shear signal that can maximally symmetrize the PDF of the shear estimators with respect to zero. Note that this is significantly different from the conventional way, which uses the weighted sum of the shear estimators. In this work, to measure the excess surface density, the idea is similar: for a radius bin we first obtain a PDF of the tangential shear estimator, $G_{\rm t}$, using galaxy shapes. The estimated excess surface density, $\widehat{\Delta \Sigma}$, is the value where the PDF of the following residual is symmetrized and centered around 0:
\begin{equation}
    G_{\rm t}(z_{\rm s}) - \frac{\widehat{\Delta \Sigma}}{\Sigma_{\rm c}(z_{\rm l},z_{\rm s})} \cdot (N + U_{\rm t})(z_{\rm s}),
    \label{eq:PDFSymm1}
\end{equation}
in which $z_{\rm l}$ and $z_{\rm s}$ are the redshifts of the lens and source respectively, and $N$ and $U_{\rm t}$ are the other components of the shear estimators defined in FQ to properly take into account the normalization and parity symmetry of $G_{\rm t}$. Equation~\ref{eq:PDFSymm1} stems from the well known relation between the excess surface density of the lens and the tangential shear signal of the background: $\Delta \Sigma(r) = \gamma_{\rm t}(r) \, \Sigma_{\rm c}$. Here we assume the cosmology given in Section~\ref{sec:simulation}. 

The errors on the $\widehat{\Delta \Sigma}$ are estimated by using 200 jackknife samples. Our discussion of the potential systematic uncertainties due to, e.g., photo-z biases, dilution by member galaxies, magnification bias, etc. are given in a separate work (Li et al., in preparation).

\subsection{Lens catalog}
\label{sec:lensCatalog}
The lens catalog used in this work is provided by the extended halo-based group/cluster finder detailed in~\cite{Yang2021}, originally developed in~\citet{Yang2005b,Yang2007}. 
The method first identifies each galaxy as a halo candidate then, weighted by their luminosities, it iteratively combines haloes into larger ones, if a halo resides within another's boundary. Finally the haloes are ranked according to the total luminosities of their member galaxies, and matched with the halo mass function prediction of a given cosmology to obtain the masses. 
The extended method can be applied to both photometric and spectroscopic redshift surveys to provide better performance for the group completeness than other photometric group/cluster finders, and is applied to the Dark Energy Spectroscopic Instrument (DESI) Legacy Image Surveys DR9 galaxy catalog. For more details please see~\cite{Yang2021}.

For this work we select haloes in the redshift range $0.2 \leq z \leq 0.3$ with masses between $12 < \log ( M_{\rm grp} ~[{\rm M_{\odot}}/h] ) \leq 15.3$, where $M_{\rm grp}$ is the halo mass enclosed within $r_{\rm grp}$, defined in the halo-based group/cluster catalog. 
We exclude haloes with $\log M_{\rm grp} \leq 12$ as these haloes have lower than $90\%$ group purity percentage~\citep{Yang2021}.
Here $M_{\rm grp}$ is the estimated halo mass from abundance matching which is provided in the group catalog. It corresponds to $M_{\rm 180m}$, the mass enclosed within $r_{\rm 180m}$ that encloses a mean density equal to $180$ times the mean matter density of the Universe, or $M_{\rm 180m}=\frac{4 \pi}{3} r_{\rm 180m}^{3} \, 180 \rho_{\rm m}$. In this work we only use $M_{\rm grp}$ for binning haloes into different mass ranges. Once the weak lensing profile is obtained from the FQ pipeline, we also derive a corresponding virial mass, $M_{\rm WL}$, by fitting the lensing profile with a 2-halo Navarro, Frenk, and White (NFW) model~\citep{NFW97} profile (see Appendix~\ref{sec:WLMassModel}). We show a comparison between the lensing mass and the group mass in Appendix~\ref{sec:massComparison}. We will use $M_{\rm WL}$ when comparing our results with simulations.

We determined the average density profile of massive objects by averaging the weak lensing signal over many haloes, a process called stacking. This boosts the weak lensing SNR from individual haloes, which usually have very small SNRs. 
In Section~\ref{sec:bin1d} we divide the haloes into five evenly spaced $\log M_{\rm grp}$ bins. 

We also study how a secondary halo parameter impacts on our results. To this end we bin the group sample in both mass and concentration.
However, measuring a concentration parameter for individual haloes with precision can be extremely difficult. For example, the concentration is traditionally $c_{\rm mdef}=r_{\rm mdef}/r_{\rm s}$, where $r_{\rm mdef}$ is the mass definition (e.g. sp, $200c$, $500m$, vir, etc.) and $r_{\rm s}$ is the scale radius located where the density profile has a slope of $-2$. A small curvature difference in the density profile can lead to a large offset in the estimation of $r_{\rm s}$. Furthermore, an incorrect halo center can underestimate the density in the central region of the halo and give lower concentration predictions~\citep{Prada2012}.

An interpretation of the concentration is that it represents the distance between two different halo scales, where the scales represents unique features of the halo, and that the concentration is typically $\geq 1$. 
With this in mind we hope to use the distribution of member galaxies as a proxy for concentration when binning our haloes.
Here we define a member galaxy concentration as $c=r_{\rm max}/r_{\rm e}$, where $r_{\rm max}$ is the distance of the farthest member galaxy from the halo center and $r_{\rm e}$ is the effective radius, also called the half-light radius which encloses half of the total luminosity of the group.
Our choice of $r_{\rm e}$ is the galaxy position (from the halo center) where the enclosed galaxy luminosity is closest to half of the total luminosity.
Most haloes in the catalog have a richness less than three, which we exclude from our analysis to avoid an overwhelming number of $r_{\rm e} = 0$ (or $c=\infty$) haloes, as the central-most galaxy is typically the brightest. 
This method will lead to coarse estimates of the effective radius, especially for the low-richness haloes. 
Because we cannot obtain precise concentration measurements, we will take the upper and lower $25\%$ of the concentrations to study the overall trend of the concentration on splashback and depletion scales. The trend should reflect those found in~\cite{Fong2021}, if the changes between $z \sim 0.2 - 0.3$ to today are smooth and monotonic on these scales.
In Section~\ref{sec:bin2d} we bin the haloes into four mass bins and then in each mass bin by the upper and lower concentration quartiles.

\section{Simulation and the halo sample}
\label{sec:simulation} 
We use the CosmicGrowth Simulations~\citep{Jing2019} to compare with our observation results. The CosmicGrowth Simulations is a grid of high resolution $N$-body simulations run in different cosmologies using a P$^3$M code~\citep{Jing2002}. We use the $\Lambda$CDM simulation with cosmological parameters $\Omega_{\rm m}=0.268$, $\Omega_{\Lambda}=0.732$, and $\sigma_{8}=0.831$. The box size is $600~{\rm Mpc}/h$ with $3072^{3}$ dark matter particles and softening length $\eta=0.01~{\rm Mpc}/h$. 
Groups are identified with the Friends-of-Friends (FoF) algorithm with a linking length $0.2$ times the mean particle separation. The haloes are then processed with \textsc{HBT+}\footnote{\url{https://github.com/Kambrian/HBTplus}}~\citep{Han2012,Han2018} to obtain subhaloes and their evolution histories. The resulting halo catalog has about $2\times10^{6}$ distinct haloes with masses $10^{11.5}< M_{\rm vir} ~[{\rm M_{\odot}}/h] < 10^{15.5}$, where the minimum mass corresponds to roughly $500$ particles within the virial radius to keep a reliable resolution on the structure of haloes. 
The virial mass, $M_{\rm vir}$, is the mass within a spherical volume of radius $\rvir$ that encloses the mean density $\Delta_{\rm c}$ times the critical density of the Universe, or $M_{\rm vir} = \frac{4\pi}{3} \rvir^{3} \, \Delta_{\rm c} \rho_{\rm crit}$. The virial overdensity in units of the critical density, $\Delta_{\rm c}$, is predicted from the spherical collapse model~\citep{Bryan1998}. Note the total mass is used in the computation of the virial mass and radius, not just the bound mass, and that the halo centres are located at the most bound particle of the central subhalo.

In previous works~\citep{Han2019,Fong2021} the halo bias was studied with many halo properties at $z=0$ using the CosmicGrowth Simulations. For our simulation predictions we will focus on haloes binned by the virial mass for $0.2 \leq z \leq 0.3$, though we will make references to the $z=0$ case and other halo properties in the results. 
We extract the individual overdensity profiles from the simulation at redshift snapshots $z = 0.289, 0.253, 0.218$, to compare with the redshifts of the lens catalog. Though we have haloes at discrete redshifts, we can obtain the general trend in the lensing profiles and radii. As we will show later, $\rcd$ evolves slowly enough for these redshifts so that we can use any of these redshifts as our comparison.

\section{The halo bias profile in weak lensing}
\label{sec:biasAndWeakLensing}
In this section we will discuss the halo bias profile and the characteristic depletion radius used in this work, and how they can be estimated using weak lensing.

\subsection{The halo bias profile and the characteristic depletion radius}
\label{sec:binnedBias}
The bias profile is a measure of the matter clustering around a halo relative to the clustering around a random particle in the Universe. 
It is defined as the ratio between the halo-matter and the matter-matter correlation functions, i.e., 
\begin{equation}
	b(r) = \frac{ \xi_{\rm hm}(r) }{ \xi_{\rm mm}(r) } = \frac{\langle \delta(r) \rangle}{ \xi_{\rm mm}(r) },
	\label{eq:binnedBias}
\end{equation}
where $\xi_{\rm hm}$ and $\xi_{\rm mm}$ are the halo-matter and matter-matter correlation functions, $\delta(r) = \rho(r)/\rho_{\rm m} - 1$ is the overdensity profile of matter around a halo, $\rho_{\rm m}$ is the mean matter density of the Universe, and the averaging is over all the haloes in each halo property bin. For an individual halo this is simply the overdensity profile of matter around the halo center divided by the underlying overdensity of the Universe. The impact of the background clustering of the Universe is removed once scaled by the matter-matter correlation function, leaving only the relevant halo profiles.
For example, on large, linear-scales, the clustering of matter around haloes is a constant bias factor times the average matter clustering around a random matter particle. 
Note that throughout this work we use univariate spline fitting onto $\xi_{\rm mm}$ from simulations at the given redshifts in the denominator, for both simulation and observations.

The binned bias is well described by~\citet{Fong2021}:
\begin{equation}
    b^{\rm{Fit}}(r)=\frac{  1+\left(\frac{r}{r_{0}}\right)^{-(  \alpha+\beta  )}  }{  1+\left(\frac{r}{r_{1}}\right)^{-(  \beta+\gamma)  }  } \times \left(  b_{0}+\left(\frac{r}{r_{2}}\right)^{-\gamma}  \right).
	\label{eq:bFit}
\end{equation}
with $r_0<r_1<r_2$. The various forms of the binned bias can be seen in~\cite{Fong2021}. An example of the bias binned by mass can be seen in the bottom panel of Figure~\ref{fig:excessSurfaceDensity_fit}. This function has four components describing the inner-most one-halo profile ($b\propto r^{-\alpha}$ for $r \ll r_0$) before the trough, the rise beyond the trough ($b\sim r^\beta$ for $r_0<r<r_1$), the decrease after the rise ($b\sim r^{-\gamma}$ for $r>r_1$) and the large-scale linear bias ($b\sim b_0$ for $r\gg r_2$). 
Note that this fitting function asymptotes to a single power-law near the center of the halo, which may not be accurate enough for describing the innermost profile in simulations, so we will generally limit our radial bins to start from $r \geq 0.06 \, {\rm Mpc}/h$. For discussions on how this is dealt with in weak lensing see Appendix~\ref{sec:WLCode}.
Regardless, we focus on studying the outskirts of haloes on intermediate to large scales, so the inner-most region is less important in this work.

In almost all cases the bias profile has a trough where the correlation between matter and haloes is the weakest, relative to the average clustering of matter in the Universe. 
In Figure~\ref{fig:excessSurfaceDensity_fit}, the bias profiles for high-mass haloes have no troughs and for decreasing masses the trough deepens. Furthermore, the trough locations decrease with decreasing mass. The formation of the trough can be interpreted in one of or a combination of two factors. In the context of halo modelling, high-mass haloes cover a larger surface area and lower-mass haloes can be packed in their environments, having a smooth transition between the halo and their environments.

In the context of halo accretion, the gravitational influence of high-mass haloes reach out on larger scales than low-mass ones. 
In~\citet{Fong2021}, the non-linear halo bias, mass, radial velocity, and mass-flow-rate (MFR) profiles are explored to build a physical picture of the trough discussed above, on physical scales. As the MFR we studied is defined in physical units, the density evolution we discuss evolves in physical space and comes from a differential net flow into a physical boundary. While haloes are growing, matter on their outskirts experiences a competition between gravity, causing material to inflow, and the expansion of the Universe as well as the gravity of the neighbouring haloes, causing matter to outflow.
Similar in shape to the radial velocity profile in Figure~\ref{fig:cartoon_depletionRegion}, the MFR profile has a clear maximum inflow rate. This location is where particles are inflowing most quickly on average. Due to continuity, inside the maximum inflow rate the halo is growing, while outside matter is being depleted. Thus, the maximum inflow rate location can be thought of as the inner depletion radius, $\rid$, the inner boundary to the region where mass is being depleted. 
When haloes are binned by mass, the MFR and radial velocity profiles have very similar shapes, and thus $\rid$ can be identified as either the maximum inflow or infall rate locations, identified from the MFR or radial velocity profiles respectively. The outer boundary of the region of depletion is the turnaround radius, $\rta$, marking the location where particles, on average, are overcome by Hubble Flow. The lowest mass haloes have typically reached their final turnaround radius, while the high mass haloes still have evolving turnaround radii~\citep{Tanoglidis2015}. 
In the context of the mass flow rate or radial velocities, low-mass haloes have small depletion regions and are more likely nearing their end of their accretion phase, while high-mass haloes have larger depletion regions and are more likely in their fast accretion phase~\citep{Fong2021}.
With this picture in mind, the low-mass haloes, which are smaller and typically older, have drained matter from their depletion region more so than their high-mass counterparts, relative to the underlying overdensity of the Universe. 
Thus, relatively lower-mass haloes have deeper troughs in their bias and relatively higher-mass haloes deplete matter on larger scales.

The above interpretation is the motivation for defining the characteristic depletion radius as the minimum of the bias which acts as a natural boundary for haloes. 
For cluster-mass haloes ($\geq 10^{14} ~{\rm M_{\odot}}/h$) we do not find a trough in the bias and $\rcd$ is estimated as the location where the bias transitions to the linear bias. For more details on the physical picture of the bias profile and the characteristic depletion radius please see~\cite{Fong2021}.

\subsection{From bias profile to lensing profile}
\label{sec:weaklensing} 
With the purpose of obtaining the bias profiles and $\rcd$ from observations, we take a forward-modelling approach and use the bias profile, Equation~\ref{eq:bFit}, to predict a corresponding excess surface mass density profile which will be subsequently fitted to the observations. 
This method assumes spherically symmetric densities, which is a reasonable assumption when stacking randomly oriented haloes together.

Using the relation in Equation~\ref{eq:binnedBias}, we can use our analytic bias profile (Equation~\ref{eq:bFit}) to obtain the spherically symmetric 3D mass density profile
\begin{equation}
\rho(r) = \rho_{\rm m} \times ( \xi_{\rm mm}(r) \times b(r) + 1).
\end{equation}
With a spherically symmetric 3D mass density profile, we obtain the 2D surface mass density by integrating along the line-of-sight, $dz$:
\begin{equation}
    \Sigma(R) = 2 \int_{0}^{\infty}\rho(R,z)dz 
    = 2\int_{R}^{\infty }{\frac {\rho(r) \, r}{\sqrt {r^{2}-R^{2}}}} \, dr,
\label{eq:Sigma}
\end{equation}
where $R$ is the projected radius relative to the center of the lens on the lens plane. The second equality in Equation~\ref{eq:Sigma} uses the Abel transform, an integral transform for spherically symmetric functions.
The mean surface mass density of the halo is given by 
\begin{equation}
    \overline{\Sigma}(R)= \frac{2}{R^2} \int_0^R R' 
    \Sigma(R') dR'.
\label{eq:meanSigma}
\end{equation}
Finally, for a spherically symmetric lens the excess surface density is given by:
\begin{equation}
    \Delta \Sigma(R) = \overline{\Sigma}(R) - \Sigma(R).
\label{eq:DeltaSigma}
\end{equation} 
Note the above equations can be applied to spherically symmetric as well spherically averaged but asymmetric lenses.

We build a module based on the bias profile (Equation~\ref{eq:bFit}) to fit the bias, 3D density, surface density, and excess surface density profiles found in this section. This code is tested extensively against the simulated haloes in Section~\ref{sec:simulation}. For some details on the code please see Appendix~\ref{sec:WLCode}.

\section{Results}
\label{sec:results} 

In this section we will bin the haloes by mass, then by mass and concentration as discussed in Section~\ref{sec:observations}. With these measurements we further discuss the potential to use the ratio between the characteristic depletion to splashback radii to understand halo accretion phase.

With Equations~\eqref{eq:bFit}-\eqref{eq:DeltaSigma}, we can fit the measured excess surface density profile to obtain the bias profile and its uncertainty, from which we can further derive the various halo radii. In this work the middle simulation redshift $z=0.253$ is used for the weak lensing calculations.
We use the affine-invariant ensemble sampler for Markov Chain Monte Carlo (MCMC) package \textsc{emcee}~\citep{emcee} and fit over the excess surface density profiles and error covariance matrices (Section~\ref{sec:observations}) to obtain the posterior samples for the bias parameters. From these samples we take out the burn-in steps and obtain $\rcd$, $\rsp$, and corresponding $\Delta$ posteriors from the bias parameters corresponding to each posterior step.
The characteristic depletion radius is determined by taking the minimum of the bias profile; the splashback radius is taken as the minimum of the logarithmic density slope; and the scaled enclosed density is taken as the enclosed density, within a given radius, scaled by the mean density of the Universe. This process gives us a posterior distribution for each of these measurements, and the values and error bars used in the results are determined by the medians and $\pm 34$ percentiles from the medians.

\subsection{Haloes binned by mass}
\label{sec:bin1d}

\begin{figure*} 
	\includegraphics[width=\textwidth]{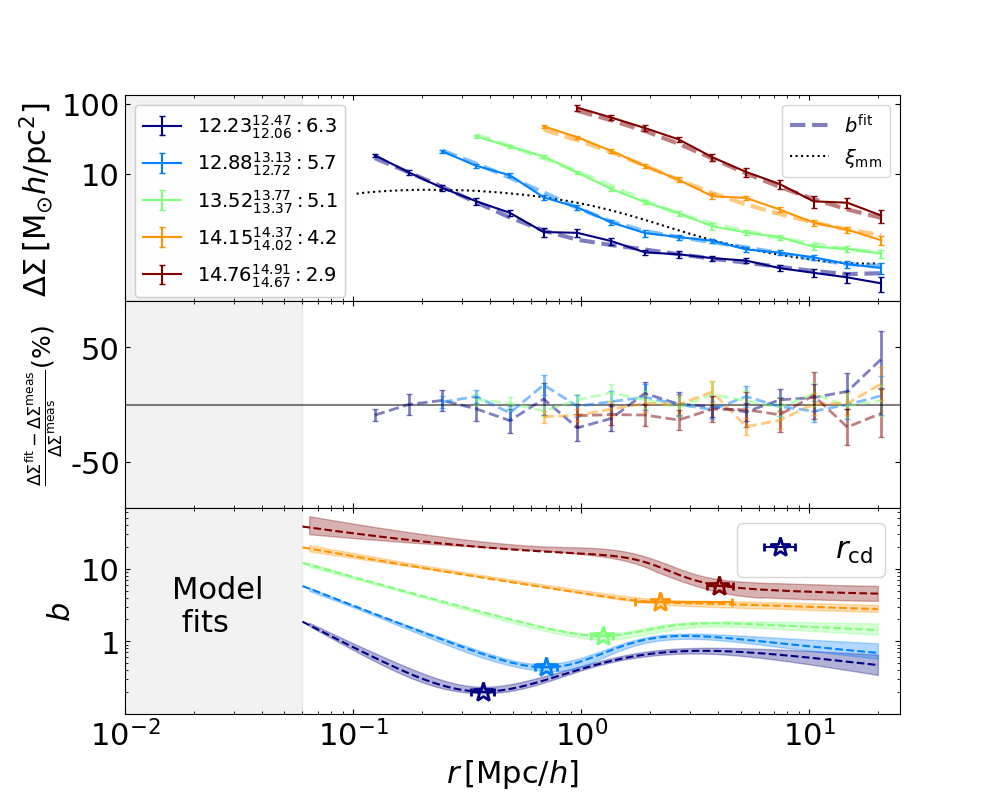}
	\caption{
	Resulting fits from MCMC for haloes binned by virial mass. The observed excess surface density and their fits, solid lines with error bars and dashed lines respectively, are shown in the top panel. The colours correspond to the median logarithmic group masses $\log M_{\rm grp}$ in the original group catalog (with upper and lower $\pm 34\%$) and the logarithm of the number of haloes in the bin, labelled in the upper left legend. We use a fit range of $R \in [0.5 \, r_{\rm grp},20] \, {\rm Mpc}/h$. 
	The fit excess surface density results, modelled on the analytic bias profile (Equation~\ref{eq:bFit}), are shown as the dashed lines. For reference, we include the excess surface density predicted from the matter-matter correlation function as the dotted line.
	The percent differences between the data and fits are shown in the middle panel. 
	The corresponding fitted halo bias profiles are shown in the bottom panel. Stars with error bars are the median (with upper and lower $\pm 34\%$ from the median) characteristic depletion radii, identified as the minima of the bias profiles. The fill between curves represent the $\pm 34\%$ of the bias profiles from the median profile for each radial bin. Note that the median bias profile is not the same as the bias profile based on the medians of the posteriors. The horizontal filled area is $r<0.06 {\rm Mpc}/h$.}
	\label{fig:excessSurfaceDensity_fit}
\end{figure*}
In the top panel of Figure~\ref{fig:excessSurfaceDensity_fit} we plot the observed excess surface densities binned according to mass and their fits, obtained by forward modeling using the bias profile (Equation~\ref{eq:bFit}),  
represented as solid lines with error bars and dashed lines, respectively. The errors in the top panel are the diagonals of the covariance matrices, but the full covariance matrices are used for the fits. Each color corresponds to a bin labelled by the median group catalog mass, $\log M_{\rm grp}$, with $\pm 34\%$ from the median and log number of lenses in each bin.
The percent difference between the data and fits are shown in the middle panel. 

In the bottom panel we also include the fitted bias profile and their corresponding characteristic depletion radii with error bars. The bias profiles use the median of the posteriors from MCMC fits to the excess surface density, while the characteristic depletion radii are the median with upper and lower $\pm 34\%$ from the $\rcd$ posteriors. The characteristic depletion radius is defined at the bias minimum, and when there is no minimum it is located where the 1-halo term transitions to the large-scale linear bias. Note that haloes with masses $\gtrsim 10^{14} ~{\rm M_{\odot}}/h$ are typically considered as clusters. This is an interesting mass scale, which roughly separates haloes that have troughs in the bias profile and those that do not~\citep{Fong2021}. In the $\sim 10^{14} ~{\rm M_{\odot}}/h$ case the bias profile does not have a steep change at the 1-halo to 2-halo transition, and so small changes in the bias profile can lead to very different transition locations. Thus small changes in the bias parameters (from the MCMC posteriors) cause a wider range in the estimated characteristic depletion radius, which accounts for the large error bars in the results.

Note that for the bias fitting we exclude the inner-most region from our fits, out to $0.5 \, r_{\rm grp}$, where $r_{\rm grp}$ is determined from $M_{\rm grp}$ assuming a NFW profile and default mass-concentration relation, using \textsc{colossus}. We coarsely vary this factor and find that our fit results perform the best for $0.5$, keeping in mind the percent differences in the middle panel. This is done to avoid systematics such as mis-centering and photometric redshift contamination that have large impacts on the central region of haloes~\citep{Han2015}. 
Because the very central region is not properly modelled in the bias profile (see Section~\ref{sec:biasAndWeakLensing}), we believe that including mis-centering into our fitting will not be very satisfactory anyway. Furthermore, we do not expect mis-centering to impact on splashback scales~\citep{More2016}, and thus less so on depletion scales.

As the bias fitting is not accurate in the inner part, to estimate a virial mass from weak lensing we choose to use a conventional halo model to fit the weak lensing mass profile. The halo model parameters includes mass, concentration, mis-centering, and the large-scale bias, so we fit a larger range ($R \in [0.02,43] {\rm Mpc}/h$) of the excess surface density (For more details see Section~\ref{sec:WLMassModel}). 
Thanks to the large number of lenses ($\sim 10^3-10^6$ in each bin) and the high source densities ($\sim 10/{\rm arcmin}^2$), we are able to get a typical statistical error on the estimated lensing mass as small as $\sim 0.02$ dex. 
To compare the weak lensing masses with the catalog masses, we map the $M_{\rm grp}$ from the cosmology used in~\cite{Yang2021} to the cosmology used in this work by matching the peak height between the two cosmologies. 
Then we convert the mass to the virial mass assuming a NFW profile and default mass-concentration relation. 
We find that the ratio between weak lensing versus halo catalog virial masses is roughly $0.96$ (see Appendix~\ref{sec:massComparison}).
This difference might be expected as each mass estimator has its own biases~\citep{Han2015, Li2021_Qingyang} and obtaining unbiased halo mass is the goal of many fields of astrophysics and cosmology. However, measuring the true halo mass is notoriously difficult (See more discussion in Section~\ref{sec:systematics}), and even though a highly precise mass measurement is ideal, it is not critical for this work.

\begin{figure} 
	\includegraphics[width=\columnwidth]{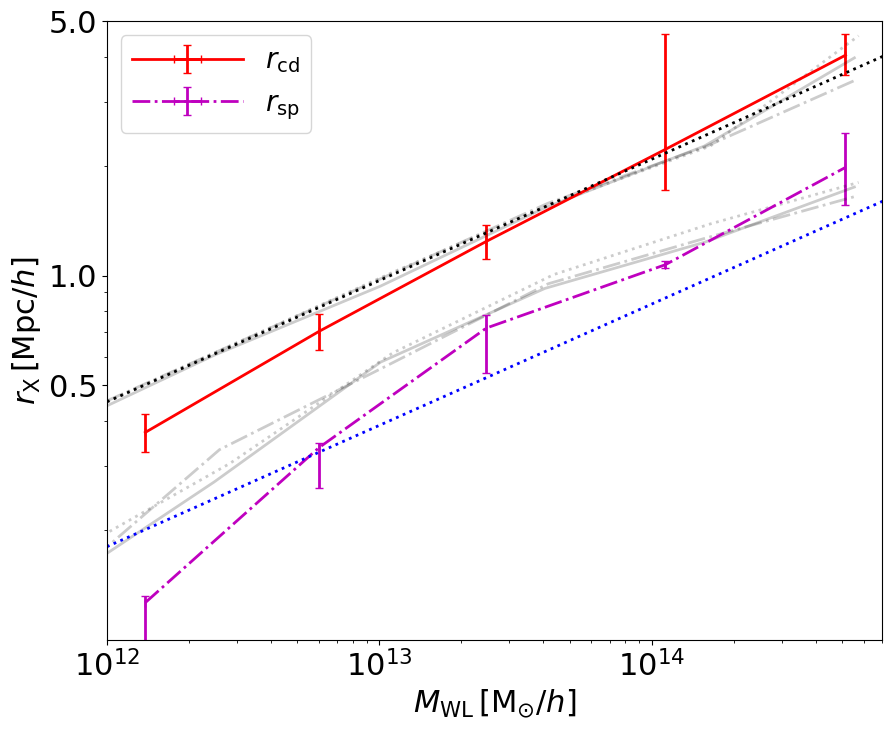}
	\caption{
	The characteristic depletion and splashback radii versus weak lensing virial mass, plotted as red solid and magenta dash-dotted lines, respectively. The virial radii are plotted as the blue dotted line as reference, with $z=0.253$. The $\rcd$ ($\rsp$) predictions from simulation snapshots $z = 0.289, 0.253, 0.218$ are represented by the upper (lower) faint solid, dot-dashed, and dotted lines, respectively. Note that simulations are plotted against their binned true virial masses. Vertical error bars are the resulting median and $\pm 34 \%$ from the median taken from the MCMC samplers, for the respective radius. Horizontal error bars are the virial mass and errors ($\sim 0.02$ dex and barely visible on the plot) from fitting the 2-halo NFW profiles onto the excess surface density profiles, discussed in the text. The dotted black line is $2.5 \, \rvir$.
	}
	\label{fig:RxVsMvir}
\end{figure}
The measured characteristic depletion and splashback radii versus weak lensing virial masses are plotted in Figure~\ref{fig:RxVsMvir}, as red solid and magenta dash-dotted lines with error bars. The simulation results at three different redshifts are also plotted as grey lines.
The characteristic depletion and splashback radii do not seem to have any obvious growth for this redshift range. As a result, in subsequent plots we will only compare our measurements with the simulation results at $z=0.253$ for clarity. A detailed study of how the depletion radii as well as the bias and density profiles evolve over time will be studied in a separate work (Gao et al., in prep.). 

We also include $2.5 \, \rvir$ as the black dotted line. In the case of simulations, $\rcd$ is in agreement with $2.5 \, \rvir$, though not as precisely as $z=0$ haloes~\citep{Fong2021}. This implies that the virial radius can be a good estimator for the depletion radius for haloes binned by mass.
The measured characteristic depletion radius has good agreement with $2.5 \, \rvir$ and the simulation predictions, with the exception of the lowest mass bin. Note that we discuss the error bars on the $\sim 10^{14} ~{\rm M_{\odot}}/h$ case above. 

For the lowest mass bin the measured $\rcd$ is lower than expected.
Though the group/cluster finder is very successful, the low measurement may be due to the systematics in the halo finder itself. For example, the halo mass assignment for each group has an uncertainty of about 0.2 dex at the high mass end~\citep{Yang2021}. With decreasing halo mass the uncertainty increases, and can potentially bias our results more significantly for relatively lower-mass bins. As there are considerably more low-mass haloes than high-mass ones, this uncertainty will bias the low-mass bins more significantly in the stacking process~\citep{Dong2019}, an effect often called the Eddington bias~\citep{Eddington1913}. This means that there are more opportunities that low-mass haloes may be misidentified as haloes of relatively higher masses, and the amplitude and shape of the stacked density profile will be biased towards a smaller halo shape, biasing the weak lensing mass and $\rcd$ low. 
However, the exact impact of the Eddington bias on both of these values are difficult to pin down due to the model dependent predictions in simulations for a given observable, which we will leave to a future work. 
Another possibility might be that the completeness of the halo catalog is not perfect, and the finder is more optimal in finding a subset of low-mass haloes rather than the entire population. For example, for low-mass haloes the characteristic depletion radius decreases with increasing formation time~\citep{Fong2021}. \textit{If} the halo finder is better at finding older low-mass haloes, then our $\rcd$ measurements would be biased low.

We also plot $\rsp$ from our observations and find that the measurements are roughly consistent with those found in our simulations, with the exception of the two lower mass bins. 
Previous measurements of the splashback radius tend to find lower values of $\rsp$ than simulation predictions~\citep{More2016,Chang2018,Contigiani2019}.
In addition to the possibilities above, other potential sources for the discrepancy have been proposed such as dynamical friction drag due to massive subhaloes~\citep{Adhikari2016}, different methods in obtaining the splashback radius~\citep{Xhakaj2020}, new physics~\citep{Adhikari2018,Banerjee2020}, or different methods in obtaining the halo profile~\citep{Murata2020}.
Our $M_{\rm WL} < 10^{13} ~{\rm M_{\odot}}$ mass haloes have splashback measurements significantly lower than our simulation results. This may be associated with the large scatter in the halo masses in these low mass groups.
To fully understand the splashback measurement sensitivity to the halo finder used in this work would require a deeper investigation on a similar level as done in~\cite{Murata2020}, and is outside the scope of this paper.
We should note that we avoid many systematics in a crude way by ignoring the central region of the halo. Though crude, we do not expect that mis-centering would have a significant impact on the splashback measurement~\citep{More2016}.

The characteristic depletion and splashback radii contain different halo property information. $\rcd$ represents the location where matter is being depleted for growing haloes, where gravity of the haloes of interest are in competition with gravity due to their neighbours as well as the expansion of the Universe. The matter being depleted is thus limited and the evolution in $\rcd$ can be predicted (Gao et al., in prep.), and gives information on halo growth in the context of the matter on the haloes' outskirts. $\rsp$ represents the location where matter is being built up during their first orbital apogees, or where the highest energy particles that are more recently captured reach their farthest distances. This has been shown to be correlated with the haloes' mass accretion rates~\citep{Diemer2014, Adhikari2014, More2015, Shi2016}. Thus, though related through halo mass accretion history, the characteristic depletion and splashback radii are expected to contain different halo properties.
In the $z=0$ simulation~\citep{Fong2021}, when haloes are binned by mass $\rcd/\rvir=2.5$ and $\rcd/\rsp \sim 2 - 3$, where higher mass haloes (with typically higher mass accretion rates) have $\rcd/\rsp \sim 2$. When haloes are binned by mass and a secondary parameter, $\rcd/\rsp$ and $\rcd/\rvir$ have more complicated relationships, yet $\rcd/\rsp$ has a similar range in ratios. These secondary parameters includes N-body proxies for concentration and mass accretion rate. Combined, these imply that the characteristic depletion radius contains separate information than what the splashback and virial radii already contain and can be used to probe differing halo properties. 

In the context of particle dynamics, as the accreted matter is undergoing their first orbital infall and approaching their first apogee, they can build up around the location where their radial velocities reach zero. The build up of matter on these scales will have an impact on the density profiles, out to the inner depletion radius. This can be seen in Figure~\ref{fig:cartoon_depletionRegion} in the region where radial velocities are positive and slowing down to zero. 
This build up matter corresponds to a steepening of the slope in the density profile, where $\rsp$ is traditionally defined at the steepest location. When matter is still being actively accreted and the halo is growing, the potential well deepens. For the splashback radius, this means that higher energy particles can then be captured and they can reach out to larger first orbital apogees. For the characteristic depletion radius the deepening of the potential well means that matter can be accreted out to larger scales. This corresponds to the trough in the bias profile that moves outward, and therefore $\rcd$ increases.
Thus the locations of the splashback and characteristic depletion radii are expected to contain information about the halo accretion phase and yet differing accretion information. 
As one example, an interesting subset of haloes are old, low-mass, and highly concentrated.\footnote{In~\citep{Fong2021}, the oldest low-mass haloes can be identified in the formation time versus mass panel of Figure 4, which corresponds to the highest concentration low-mass haloes. The corresponding $\rsp/\rcd$ ratios can be seen in Figure 6.}
These haloes, which have mostly completed their accretion phase after being in the proximity of massive neighbours~\citep{Wang2007}, have the highest $\rsp/\rcd$ ratios.
Furthermore, the higher mass haloes which typically younger and are in their early accretion phase, have relatively lower $\rsp/\rcd$ ratios.
This trend implies that $\rsp/\rcd \approx 2$ when haloes are in their fast accretion phase, while $\rsp/\rcd \approx 3$ in their slow accretion phase. 
This exercise can be applied to different halo samples in simulations, and once other observational secondary halo proxies are discovered, can be compared with each other as tests.
Thus these measurable radii in combination can be used to probe halo properties beyond what an individual radius can and can potentially be used to test the non-linear regime of cosmologies in simulations.

\begin{figure} 
	\includegraphics[width=\columnwidth]{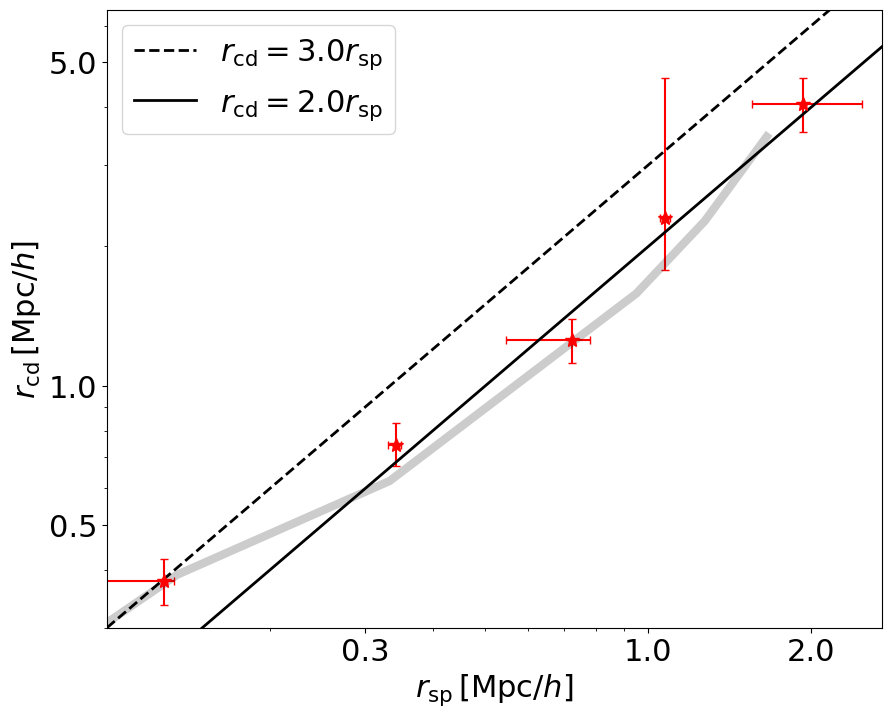}
	\caption{
	The characteristic depletion versus splashback radii. The vertical and horizontal error bars are the median and $\pm 34 \%$ from the median, for the characteristic depletion and splashback radii inferred from the lensing profiles, respectively. The dashed and solid lines are the ratios $\rcd / \rsp=3.0$ and $2.0$, respectively. Simulation predictions are represented by the faint grey line.
	}
	\label{fig:RcdVsRsp}
\end{figure}
In Figure~\ref{fig:RcdVsRsp} we plot the relation between $\rcd$ and $\rsp$ directly, and compare the measured results to those found in simulations. 
The measurements and simulation expectations agree throughout the entire mass range, even in the lowest-mass bins.
We include the ratios $\rcd / \rsp=3.0$ and $2.0$ represented as dashed and solid lines, to compare with the $z=0$ case in~\cite{Fong2021}. 
In simulations haloes with virial masses $\leq 10^{12} ~{\rm M_{\odot}}/h$ typically have an outer concentration, $\rcd/\rsp$, of 3, and higher mass haloes have lower outer concentrations, having a minimum outer concentration of roughly $\sim 2$ for intermediate masses. 
The confirmation of the measured outer concentration, $\rcd/\rsp$, has very interesting potential uses in studying the halo profile. This, along with $\rcd/\rvir=2.5$, implies an approximately universal outer halo profile, and the deviations from universality, in the form of the outer concentration ($\rcd/\rsp$), is likely due to accretion history~\citep{Fong2021}.

\begin{figure} 
	\includegraphics[width=\columnwidth]{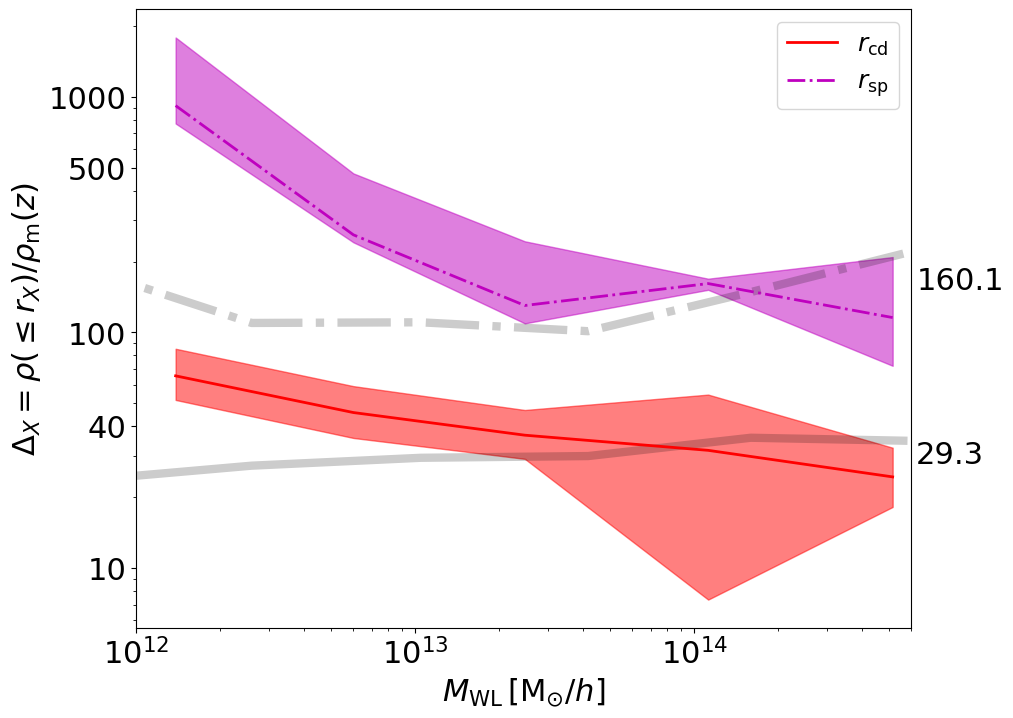}
	\caption{
	The average density enclosed in different radii, $r_{X}$ as labelled in the legend, scaled by the mean density of the Universe at $z$. The bottom red solid (top magenta dash-dotted) line represents the scaled enclosed density within the measured characteristic depletion (splashback) radius. The simulated $z=0.253$ counterpart is represented as the grey line for each. The values on the right are horizontal fits to the observations, excluding the two lowest mass bins.
	}
	\label{fig:DeltaxVsMvir}
\end{figure}
In Figure~\ref{fig:DeltaxVsMvir} we plot the enclosed density within the radii, $r_{X}$ labelled in the legend, normalized by the mean density of the Universe at $z$. Here we use the relation $M_{X}=\frac{4 \pi}{3} \Delta_{X} \rho_{\rm m}(z) r_{X}^{3}$, where $M_{X}=M(\leq r_{X})$ is the mass enclosed within $r_{X}$ and the normalized enclosed density is $\Delta_{X}=\rho(\leq r_{X}) / \rho_{\rm m}(z)$. 
$M_{X}$ is obtained by interpolating the enclosed mass, from the bias fit, at the location of $r_{X}$.\footnote{Note that, though the bias fit model used in this work (Equation~\ref{eq:bFit}) is not optimized for the very central region of the bias in simulations (Section~\ref{sec:biasAndWeakLensing}), this has negligible impact for the simulation results, and so we suspect similarly for the observation results.}
Here the characteristic depletion and splashback radii from observations and their $\pm 34 \%$ from the medians are plotted as red and magenta fill between lines.
We fit horizontal lines to the median $\Delta$ values for both $\rcd$ and $\rsp$ weighted by their standard deviations, excluding the two lowest mass bins, obtaining $\Delta_{\rm cd}=29\pm 15$ and $\Delta_{\rm sp}=160\pm 25$. The predictions from simulations are plotted as faint lines for both radii.
Overall we see a rough agreement for halo masses $\geq 10^{13} ~{\rm M_{\odot}}$ between the observed and simulated $\Delta_{\rm cd}$, though with very large error bars. For example, the horizontal fit line to the simulated $z=0.253$ case is $\Delta=33$. 

Note the simulation results at $z=0.253$ show a lower enclosed depletion density than that found in \cite{Fong2021} at $z=0$. A complete study of this redshift evolution in simulation will be presented in an upcoming work (Gao et al. in prep.).

\subsection{Haloes binned by mass and concentration}
\label{sec:bin2d}
In this section we study how halo concentration impacts on the halo profile on characteristic depletion and splashback scales. 
In observations we start by binning the haloes by mass, and then further bin each mass bin by the upper and lower concentration quartiles, $c_{\rm high}$ and $c_{\rm low}$, respectively (See~\ref{sec:observations} for details). Then we follow the steps given at the beginning of Section~\ref{sec:results} to obtain the radii and masses studied in this work.

We also compare our measurements to the simulated $z=0.253$ case by following the same binning method above, but using $V_{\rm max}/V_{\rm vir}$ as our proxy for concentration, where $V_{\rm max}$ is the maximum circular velocity and $V_{\rm vir}$ is the circular velocity at the virial radius, and their ratio is positively correlated with halo concentration~\citep{Gao2007,Angulo2008,Sunayama2016}.
It is important to point out that in this paper $V_{\rm max}$ is calculated using only particles bound to the main subhalo, excluding satellite and unbound particles. When $V_{\rm max}$ is determined using all the particles around the halo center, the difference between high-$V_{\rm max}/V_{\rm vir}$ and low-$V_{\rm max}/V_{\rm vir}$ is much smaller. This brings into question of which $V_{\rm max}$ measurement we should consider in simulations and which are more representative of the luminosity-weighted galaxy membership distribution concentration proxy used in this work. However, in this paper we are using N-body simulations that do not allow for a direct comparison with the galaxy distribution, so we will leave this investigation to a future work.
Due to the above, we cannot make any strong claims when we compare the observational results to simulation results, though we will leave some discussion in Appendix~\ref{sec:VmaxDiscussion}. However, we will plot the simulation results when $V_{\rm max}$ is calculated using only the central subhalo, where the difference between high- and low-$V_{\rm max}/V_{\rm vir}$ is more significant.

\begin{figure}
     \centering
         \centering
         \includegraphics[width=\columnwidth]{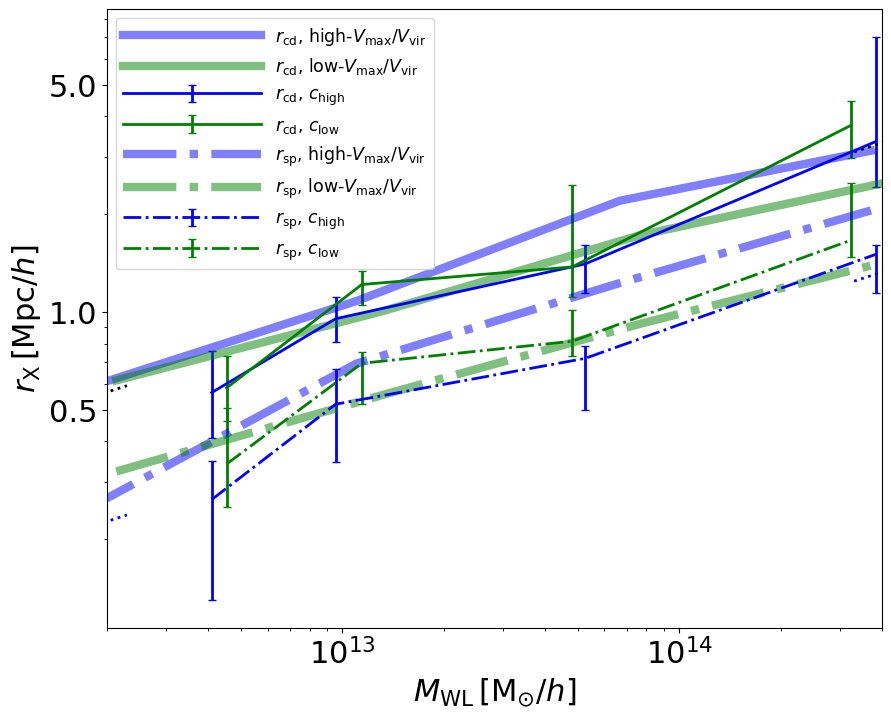}
         \label{fig:splitVmax_RxVsMvir}
        \caption{
        The measured characteristic depletion and splashback radii binned by mass and concentration. The measurements are compared with N-body simulation when haloes are binned by mass and concentration proxy. 
        Similar to Figure~\ref{fig:RxVsMvir}, but with the high (low) concentration quartiles, $c_{\rm high}$ ($c_{\rm low}$), plotted in blue (green). The simulation results for $z=0.253$ for high and low $V_{\rm max}/V_{\rm vir}$ are represented by the thick transparent blue and green lines respectively. The upper solid and lower dash-dotted lines are $\rcd$ and $\rsp$, respectively. 
        }
        \label{fig:splitC_RxVsMvir} 
\end{figure}
The measured high and low concentration quartile characteristic depletion and splashback radii can be seen in Figure~\ref{fig:splitC_RxVsMvir} as blue and green data points with errorbars, respectively. The upper solid lines represent $\rcd$ while the lower dash-dotted lines represent $\rsp$. The thick transparent lines are the simulation results for $z=0.253$, where the upper solid lines represent the characteristic depletion radii, and the lower dash-dotted lines represent the splashback radii. 
For reference we include $\rvir$ and $2.5 \, \rvir$ as the blue and black dotted lines, respectively, but only at the low and high mass ends for clarity. 

For lower concentrations the measured $\rcd$ are typically only slightly larger than the higher concentrations. We also see that the measured $c_{\rm low}$ haloes have $\rsp$ (or $\rsp/\rvir$ ratios) higher than their $c_{\rm high}$ counterparts. However, we emphasize that the uncertainties in the measurements are large so that the significances of the differences should be taken as marginal. 
The splashback observation measurements are consistent with the findings of~\citet{More2016}, where relatively lower concentrations correspond to higher measured splashback-to-virial ratios than their higher concentrated counterparts for galaxy cluster mass haloes. However, we should note that their proxy for concentration is different from ours, and that the halo catalog is constructed differently than the ones used in this paper.

\begin{figure}
    \centering
    \includegraphics[width=\columnwidth]{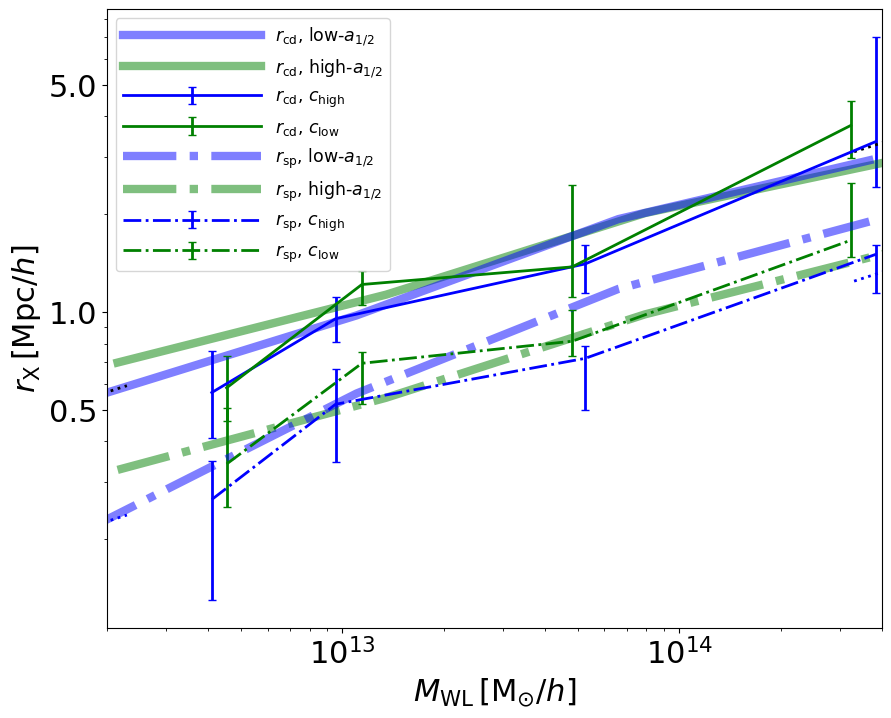}
    \caption{Same as Figure~\ref{fig:splitC_RxVsMvir}, but the simulation samples are split according to the halo formation time, $a_{1/2}$, instead of $V_{\rm max}/V_{\rm vir}$.}
    \label{fig:split_formationtime_Rx}
\end{figure}
Despite the large uncertainties, the observational measurements are qualitatively in disagreement with the simulated $z=0.253$ cases, with opposite dependences on the concentrations. This may be because the concentration proxy used in our observations may not correlate well with the $V_{\rm max}/V_{\rm vir}$ concentration proxy. In addition, the scatter of halo properties in each bin could also lead to biases in the measurements. To further assess the significance of these discrepancies, 
in Figure~\ref{fig:split_formationtime_Rx} we further compare our measurements to simulation predictions split according to halo formation time, which is expected to be a physical driver for halo concentration. As the depletion and splashback radii are also expected to be shaped by the halo formation history, splitting halos according to formation time may capture a more clear variation in these radii. The formation time, $a_{1/2}$, is defined to be the scale factor when a halo grows to half of its final mass. It is well known that $a_{1/2}$ and concentration are anti-correlated, so that old haloes are more concentrated. The results shown in Figure~\ref{fig:split_formationtime_Rx} are consistent with what we found previously using $V_{\rm max}/V_{\rm vir}$. Simulated high $a_{1/2}$ (and hence low concentration) halos have a larger $\rsp$ at the high mass end, while in observations the low concentration halos tend to have a larger $\rsp$. Interestingly, there is little dependence of $\rcd$ on $a_{1/2}$ at the high mass end in the simulation, which better matches the lack of a strong concentration dependence in observations. This indicates that the difference between our simulation and observation measurements of $\rsp$ could be more susceptible to systematics in the concentration proxy.

Keeping in mind the discussion above, this difference can be investigated using hydrodynamic simulations, which will require careful treatment of observational systematics and differing concentration proxies, which we will leave to a future work.

\begin{figure} 
	\includegraphics[width=\columnwidth]{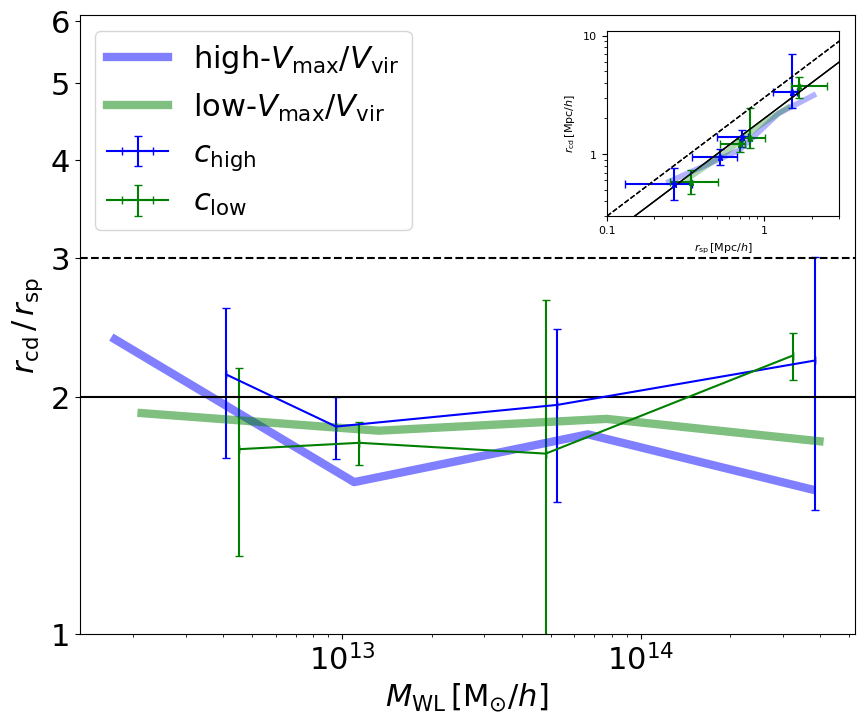}
	\caption{The ratio between characteristic depletion and splashback radii binned by mass and concentration. We show the ratio 3 and 2 represented by horizontal dashed and solid black lines, respectively. We propagate the errors for the vertical error bars. The inset subplot is similar to Figure~\ref{fig:RcdVsRsp} but for the $c_{\rm high}$ and $c_{\rm low}$ samples. The simulated high- and low-$V_{\rm max}/V_{\rm vir}$ concentration proxy quartile results are shown as the thick transparent blue and green lines, respectively.
	}
	\label{fig:splitC_RcdVsRsp}
\end{figure}
In Figure~\ref{fig:splitC_RcdVsRsp} we show the ratios $\rcd/\rsp$, or the outer concentration, for haloes binned by mass and concentration. Here the error bars are significant and any trend is difficult to discern. In the inset subplot we show that both $c_{\rm high}$ and $c_{\rm low}$ samples still tend to follow the mass dependent outer concentration relation. 
For the measured upper and lower concentrations the ratios are roughly around 2. This is in rough agreement with the $z=0$ case~\citep{Fong2021}, though there is some large disagreement for the $z=0.253$ case shown here, specifically low-$V_{\rm max}/V_{\rm vir}$ cluster-mass haloes shown here. This discrepency may be due to potential systematics (Section~\ref{sec:systematics}) or the difficulty in choosing a characteristic depletion radius for cluster-mass haloes (see Figure~\ref{fig:excessSurfaceDensity_fit}).

Note that the outer concentrations do not reach $\sim 3$, as they do in Figure~\ref{fig:RcdVsRsp}. The lowest mass bin in these cases have higher median masses ($\approx 12.5 ~{\rm M_{\odot}}/h$) due to the halo selection described in Section~\ref{sec:lensCatalog}, where we remove haloes with galaxy members of two or less and we reduce the number of mass bins to compensate for number of haloes in each bin when taking the upper and lower concentration quartiles.

For the last comparison with~\cite{Fong2021} we bin $\Delta$ by both mass and concentration. The fits to high and low concentrations for $\Delta=\rho(\leq \rcd) / \rho_{\rm m}(z)$ are $35\pm 17$ and $28\pm 50$, respectively. As the error bars are so large and there is no by-eye trend, we do not show the plot here for simplicity.

\section{A few potential systematics}
\label{sec:systematics} %
The purpose of this work is to show that the characteristic depletion radius can be measured and is consistent with simulation predictions. Though our results align well with simulation predictions, we wish to mention a few possible improvements for future measurements that we have not included in this paper.

The group/cluster halo catalog used in this work is already impressive and contains a large number of objects that covers a wide range of masses.
Even so, we remove the lower mass haloes due to lower purity and higher mass assignment uncertainties~\citep{Yang2021}. In the future, alternative halo finders that are fine-tuned to low-mass objects can potentially be combined to study the wider range of masses with higher confidence. This is particularly important for studying how the characteristic depletion and splashback radii differ for masses $\leq 12.5 ~{\rm M_{\odot}}/h$.
Furthermore, using alternative halo finders can be used to study how different halo identification tracers and methods can impact on the density profiles and the splashback and characteristic depletion radii measurements, in a similar way to~\cite{Murata2020}. 

In this work we use the NFW weak lensing masses to represent the stacked binned halo masses. Weak lensing masses are expected to be biased estimators of the true underlying masses~\citep{King2001,Becker2011,Oguri2011a,Bahe2012,Henson2017,Lee2018,Fong2018,Grandis2021}.
A few systematics of weak lensing mass bias are asphericity, dynamical state, photometric redshift uncertainties, mis-centering, halo member contamination, uncorrelated LSS, assumed density model, and most importantly the mass dispersion within the mass bin which typically leads to an overestimate in the WL mass~\citep{Han2015}. 
To calibrate weak lensing mass estimates is an important and an ongoing field.

In the FQ pipeline, we require the background source redshift to be larger than $0.2$ plus the foreground redshift. This ideally should take care of some of the systematics mentioned above as well as the impact of intrinsic alignment, one of the most important systematic effects for weak lensing~\citep{Hirata2004,Troxel2015,Yao2017,Yao2020,Shi2021}. However, spectroscopic redshifts is ideal and can improve the precision of our weak lensing measurements.

With the ongoing DESI survey~\citep{Dey2019}, the halo and source catalogs can be updated once DESI has obtained spectroscopic redshifts on the entire expected footprint. This is expected to significantly improve our redshift estimations and measurement errors of the weak lensing measurements, which can address many of the systematics mentioned above.

Many works have shown that baryonic physics and massive neutrinos can impact on the dark matter distribution of haloes and the LSS~\citep{Hennawi2005,Schaller2015,Henson2017,Mummery2017,Lee2018,McCarthy2018,Fong2018,Fong2019}. 
It is currently not understood how this will impact our simulation expectation comparisons on splashback and depletion scales, and requires a deeper investigation for high-precision measurements of the depletion radii. 

We use the luminosity-weighted galaxy member distributions as a coarse proxy for halo concentration. In our current understanding there is no precise concentration proxy for individual haloes, which is required for studying the impact of concentration in detail. For example, the mean or median of the galaxy membership separations were used as a proxy for concentration in the past~\citep{Miyatake2016,More2016}, and similar to this work it can only study the trends in concentrations. Until a better model for halo concentration comes along, our concentration trends can only be approximate.

For the mass and redshift range used in this work, about $90\%$ of the haloes is expected have membership completeness of $> 60\%$ and $85\%$ of the haloes having an interloper fraction (galaxies that belong to a different halo over the total number of true members of the halo of interest) $<1$~\citep{Yang2021}. However, because we use the luminosity-weighted galaxy member distributions as a proxy for halo concentration, this may impact our halo concentration assignments.
Another source of systematics is that there is the scatter of halo properties in each bin, potentially leading to biases in the measurements. For example, the scatter in the concentration-mass relation can partially be explained by minor and major merger events~\citep{Wang2020}. This may have drastic changes on the concentration as well as the halo density profile shape and thus relevant radii locations.

It will be very promising to use alternate observational proxies for other secondary halo parameters. For example star formation rates of galaxy members can be a tracer for halo age, and will be targets for the Subaru Prime Focus Spectrograph~\citep{Takada2014} and DESI surveys~\citep{DESI2016a,DESI2016b}. However, to improve our understanding of this relationship, we need to pin down the systematics in the methods in measuring and simulating star formation rates~\citep{Katsianis2020,Katsianis2021}.

An obvious potential improvement is with the current bias model. The bias fitting function (Equation~\ref{eq:bFit}) is a form-fitting one, and a physically motivated bias model can potentially improve our understanding of the halo bias and our estimation of the depletion radii. Ideally, the new profile would also be able to describe the very inner halo profile, and thus we can properly incorporate mis-centering into our model.

\section{Summary and Conclusions}
\label{sec:conclusions}
In this paper we use weak lensing measurements to make the first measurement of the characteristic depletion radius, $\rcd$, a newly discovered halo boundary.  
For the lenses we use the extended halo-based group/cluster finder applied to the DESI Imaging Surveys DR9~\citep{Yang2021}. For the shear measurements we use the \textsc{Fourier\_Quad} pipeline~\citep{Zhang2019} applied to the DECaLS galaxies, which provides the weak lensing profiles for this work. 
We then compare our results with a high-resolution cosmological N-body simulation from the \textsc{CosmicGrowth} simulation suit~\citep{Jing2019}.
The haloes are chosen such that $0.2 \leq z \leq 0.3$ and are binned by mass and also by mass and concentration. 
In this paper we also measure the splashback radius, $\rsp$, and virial radius, $\rvir$ to further test the theories found in~\cite{Fong2021}. When haloes are binned by mass, with the exception of the lowest mass bins, our $\rcd$ measurements show consistency with the simulation predictions. When haloes are binned by both mass and a proxy for halo concentration, we find that the depletion radius has little dependence with concentration for this mass range, and that the simulation prediction is also sensitive to the choice of concentration proxy. 
We summarize our findings below.

\textbf{Haloes binned by mass}:
\begin{itemize}
    \item The measured $\rcd$ is consistent with simulation expectations, with exception to the lowest mass bins where the halo catalog may suffer from large systematics.
    
    \item The ratio $\rcd/\rvir \sim 2.5$, and that the virial radii can be used to estimate the characteristic depletion radius when haloes are binned by mass.
    
    \item The outer concentration, $\rcd/\rsp$, follows the predicted mass-dependent behaviour $\sim 1.7 - 3$. 
    
    \item The enclosed density within $\rcd$, with exception to the lowest mass bins, is measured to be roughly $29\pm 15$ times the mean matter density of the Universe in the redshift range of our sample.
    
    \item The splashback measurements are approximately consistent with simulations for the most massive haloes. However, at lower masses the they fall below the simulation predictions, in line with previous findings.
\end{itemize}

\textbf{Haloes binned by mass and concentration}:
\begin{itemize}
    \item We do not detect a significant dependence of $\rcd$ on the concentration of galaxy distribution in our sample, for which the simulation predictions is also sensitive to the choice of the concentration proxy. 
    
    \item The observed concentration dependence on $\rsp$ is such that higher concentrations typically have lower splashback values, for the halo masses explored in this work. This trend is consistent with a previous observation~\citep{More2016}, but in the opposite direction to the expected concentration dependence in simulations. 
    
    \item Haloes with masses $\geq 10^{12.5} ~{\rm M_{\odot}}/h$, independent of concentration quartile, have outer concentrations of $\rcd/\rsp \approx 2$. 
    
\end{itemize}

When haloes are binned by mass, we find that the outer concentration has a mass-dependent behaviour. This along with $\rcd/\rvir=2.5$ describes a nearly universal halo profile, where the trend in the outer concentration is roughly $\rcd/\rsp \sim 2 - 3$ from high to low mass haloes, and can provide information on the accretion phase of haloes. As both the characteristic depletion and splashback radii can be measured in the stacked halo profiles, these measurements demonstrate the possibility of probing halo properties observationally.

Despite this possibility, it may not be trivial to find an accurate secondary proxy beside mass for selecting haloes of different depletion radii. As a proof of concept we have tried using the concentration of the luminosity-weighted galaxy distribution to bin haloes in addition to mass. Comparison with simulations suggest that splitting haloes according to the galaxy concentration of our groups may not show differing characteristic depletion radii for masses $\geq 10^{12.5} ~{\rm M_{\odot}}/h$, though do show differences for the splashback radii. Note that we expect larger concentration and formation time dependencies in the depletion radius for halo masses $\lesssim 10^{12.5} ~{\rm M_{\odot}}/h$ from simulations. Unfortunately the halo masses explored in this work are larger than this threshhold.
    
Furthermore, $\rcd/\rsp \approx 2$ for our concentration proxy in both observations and simulations. Lower masses may need to be explored to see the differences in the $\rcd/\rsp$ ratio with respect to their concentration dependencies, expected from $z=0$ simulations~\citep{Fong2021}.

Radii that describe different features of a halo can be used in combination, and can give us new and measurable insights into how the building blocks of our Universe evolve. We hope that the positive measurements of the characteristic depletion radius will open the door to new ways in studying the halo profile.

\section*{Acknowledgements}

This work is supported by the National Key Basic Research and Development Program of China (No. 2018YFA0404504), and the NSFC grants (11973032, 11673016, 11621303, 11833005, 11890692, 11890691, and 12073017), and Shanghai Natural Science Foundation,grant No.15ZR1446700 and 111 project No. B20019. We acknowledge the science research grants from the China Manned Space Project with Nos.  CMS-CSST-2021-A01, CMS-CSST-2021-A02, CMS-CSST-2021-A03. The computation of this work was done on the \textsc{Gravity} supercomputer at the Department of Astronomy, Shanghai Jiao Tong University,  and the $\pi$2.0 cluster supported by the Center for High Performance Computing at Shanghai Jiao Tong University.

The Legacy Surveys consist of three individual and complementary projects: the Dark Energy Camera Legacy Survey (DECaLS; Proposal ID \#2014B-0404; PIs: David Schlegel and Arjun Dey), the Beijing-Arizona Sky Survey (BASS; NOAO Prop. ID \#2015A-0801; PIs: Zhou Xu and Xiaohui Fan), and the Mayall z-band Legacy Survey (MzLS; Prop. ID \#2016A-0453; PI: Arjun Dey). DECaLS, BASS and MzLS together include data obtained, respectively, at the Blanco telescope, Cerro Tololo Inter-American Observatory, NSF’s NOIRLab; the Bok telescope, Steward Observatory, University of Arizona; and the Mayall telescope, Kitt Peak National Observatory, NOIRLab. The Legacy Surveys project is honored to be permitted to conduct astronomical research on Iolkam Du’ag (Kitt Peak), a mountain with particular significance to the Tohono O’odham Nation.

NOIRLab is operated by the Association of Universities for Research in Astronomy (AURA) under a cooperative agreement with the National Science Foundation.

This project used data obtained with the Dark Energy Camera (DECam), which was constructed by the Dark Energy Survey (DES) collaboration. Funding for the DES Projects has been provided by the U.S. Department of Energy, the U.S. National Science Foundation, the Ministry of Science and Education of Spain, the Science and Technology Facilities Council of the United Kingdom, the Higher Education Funding Council for England, the National Center for Supercomputing Applications at the University of Illinois at Urbana-Champaign, the Kavli Institute of Cosmological Physics at the University of Chicago, Center for Cosmology and Astro-Particle Physics at the Ohio State University, the Mitchell Institute for Fundamental Physics and Astronomy at Texas A\&M University, Financiadora de Estudos e Projetos, Fundacao Carlos Chagas Filho de Amparo, Financiadora de Estudos e Projetos, Fundacao Carlos Chagas Filho de Amparo a Pesquisa do Estado do Rio de Janeiro, Conselho Nacional de Desenvolvimento Cientifico e Tecnologico and the Ministerio da Ciencia, Tecnologia e Inovacao, the Deutsche Forschungsgemeinschaft and the Collaborating Institutions in the Dark Energy Survey. The Collaborating Institutions are Argonne National Laboratory, the University of California at Santa Cruz, the University of Cambridge, Centro de Investigaciones Energeticas, Medioambientales y Tecnologicas-Madrid, the University of Chicago, University College London, the DES-Brazil Consortium, the University of Edinburgh, the Eidgenossische Technische Hochschule (ETH) Zurich, Fermi National Accelerator Laboratory, the University of Illinois at Urbana-Champaign, the Institut de Ciencies de l’Espai (IEEC/CSIC), the Institut de Fisica d’Altes Energies, Lawrence Berkeley National Laboratory, the Ludwig Maximilians Universitat Munchen and the associated Excellence Cluster Universe, the University of Michigan, NSF’s NOIRLab, the University of Nottingham, the Ohio State University, the University of Pennsylvania, the University of Portsmouth, SLAC National Accelerator Laboratory, Stanford University, the University of Sussex, and Texas A\&M University.

BASS is a key project of the Telescope Access Program (TAP), which has been funded by the National Astronomical Observatories of China, the Chinese Academy of Sciences (the Strategic Priority Research Program “The Emergence of Cosmological Structures” Grant \# XDB09000000), and the Special Fund for Astronomy from the Ministry of Finance. The BASS is also supported by the External Cooperation Program of Chinese Academy of Sciences (Grant \# 114A11KYSB20160057), and Chinese National Natural Science Foundation (Grant \# 11433005).

The Legacy Survey team makes use of data products from the Near-Earth Object Wide-field Infrared Survey Explorer (NEOWISE), which is a project of the Jet Propulsion Laboratory/California Institute of Technology. NEOWISE is funded by the National Aeronautics and Space Administration.

The Legacy Surveys imaging of the DESI footprint is supported by the Director, Office of Science, Office of High Energy Physics of the U.S. Department of Energy under Contract No. DE-AC02-05CH1123, by the National Energy Research Scientific Computing Center, a DOE Office of Science User Facility under the same contract; and by the U.S. National Science Foundation, Division of Astronomical Sciences under Contract No. AST-0950945 to NOAO.

The Photometric Redshifts for the Legacy Surveys (PRLS) catalog used in this paper was produced thanks to funding from the U.S. Department of Energy Office of Science, Office of High Energy Physics via grant DE-SC0007914.

\section*{Data Availability}
The data underlying this article will be shared on reasonable request to the corresponding author.



\bibliographystyle{mnras}
\bibliography{ref} 




\appendix

\section{Weak lensing mass estimates}
\label{sec:WLMassModel}

\begin{figure} 
	\includegraphics[width=\columnwidth]{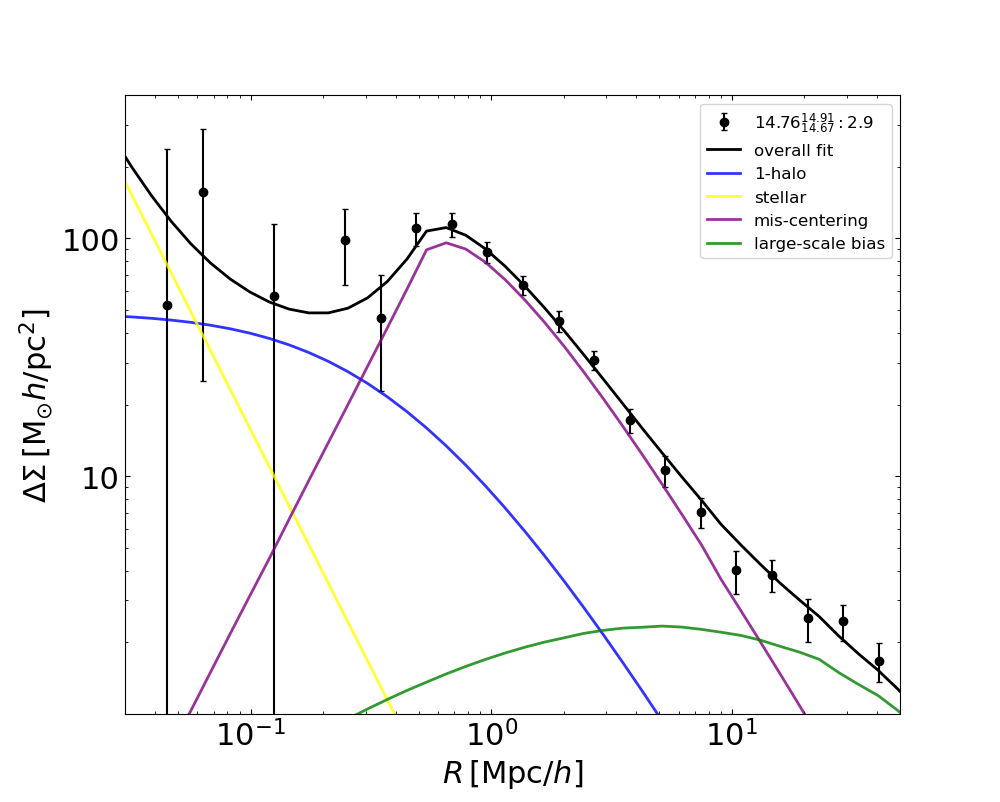}
	\caption{
	Weak lensing mass fit example for the most massive bin when haloes are binned by mass. The median and $\pm 34\%$ from the median of $\log M_{\rm grp}$ and number of haloes in the bin is in the legend. The overall weak lensing fit is represented by the black curve and its components are listed in the legend.
	}
	\label{fig:WLMassModel}
\end{figure}

The bias model in Section~\ref{sec:biasAndWeakLensing} is not optimized for modelling the very central region of haloes. To obtain the weak lensing mass, $M_{\rm WL}$, we instead use a halo model that includes both 1-halo and 2-halo terms (Wang et al., in prep.), where the model is used to obtain the mass-concentration relation and halo mass function for the same lens and source catalogs and FQ pipeline as used in this work. In this section we will briefly discuss the WL mass model. This model includes a 1-halo, stellar, mis-centering, and large-scale bias components. An example of the profile used in this work can be seen in Figure~\ref{fig:WLMassModel}.

The 1-halo term in this model is the NFW profile~\citep{NFW97}. 
The stellar component is $\frac{M_{\star}}{\pi R^{2}}$, where $M_{\star}$ is the stellar mass of the Brightest Central Galaxy (BCG), obtained by $K$-corrections~\citep[][v4\_3]{Blanton2007} applied to the BCG flux data.
To account for mis-centering we use the Rayleigh distribution proposed by~\citet{Johnston2007}:
\begin{equation}
    P_{\rm Rayleigh}\left(R_{\rm mis}\right)=\frac{R_{\rm mis}}{\sigma_{\rm mis}^{2}} \exp \left(-\frac{1}{2}\left( R_{\rm mis} / \sigma_{\rm mis}  \right)^{2}\right),
\end{equation}
where $R_{\rm mis}$ is the magnitude of the offset between the assumed halo center from the true halo center, and $\sigma_{\rm mis}$ is the width of the distribution. The former is not a free parameter, and is integrated over the distribution. To implement the mis-centering into the model we use: 
\begin{equation}
\begin{aligned}
   \Sigma^{\operatorname{\rm mis}}\left(R \mid \sigma_{\rm mis}\right)= & \frac{1}{2 \pi} \int \mathrm{d} R_{\rm mis}' P_{\rm Rayleigh}\left(R_{\rm mis }' | \sigma_{\rm mis }\right) \\
    &\int_{0}^{2 \pi} \mathrm{d} \theta \Sigma\left(\sqrt{R^{2}+R_{\rm mis}^{2}+2 R R_{\rm mis} \cos \theta}\right).
\end{aligned}
\end{equation}
The second integration is over the angle $\theta$ on the lens plane. Lastly, the large-scale bias, or two-halo term, is accounted for by~\cite{Oguri2011a,Oguri2011b}:
\begin{equation}
    \Delta \Sigma^{\rm 2h}(\theta ; M, z)=\int \frac{l \, {\rm d} l}{2 \pi} J_{2}(l \theta) \frac{\rho_{\rm m}(z) b_{\rm f}}{(1+z)^{3} D_{\rm A}^{2}(z)} P_{\rm lin}\left(k_{l} ; z\right).
\end{equation}
Here $D_{\rm A}\left(z\right)$ is the angular diameter distance at redshift z, $J_{2}(l \theta)$ is the second-order Bessel function, $P_{\rm lin}\left( k,z \right)$ is the linear power spectrum, $k_{l}=l/((1+z)D_{\rm A})$, and $b_{\rm f}$ is the large-scale linear bias factor. For the latter we use the linear bias as a function of mass $b_{\rm h}(M)$~\citep{Tinker2010} times a constant factor $f$, such that $b_{\rm f}=f \times b_{\rm h}(M)$. The overall mass model is thus given by:
\begin{equation}
\begin{aligned}
    \Delta \Sigma \left( R \right) = &
    P_{\rm cen}  \Delta \Sigma_{\rm NFW}(R) \\
    &+( 1 - P_{\rm cen} ) \, \Delta \Sigma^{\rm mis}_{\rm NFW} ( R | R_{\rm mis} ) + \Delta \Sigma^{\rm 2h} ( R ),
\label{eq:WLMassModel}
\end{aligned}
\end{equation}
where $P_{\rm cen}$ is the halo centering fraction. Overall, there are a total of $5$ free parameters: $M_{\rm WL}$ and $c_{\rm WL}$ from the NFW 1-halo profile; $\sigma_{\rm mis}$ from the mis-centering component; $f$ from the 2-halo term; $P_{\rm cen}$ from the overall model. All of the parameters are explored in Wang et al. (in prep.), though for this work we only use the weak lensing masses.

\subsection{Comparing weak lensing mass with group catalog mass}
\label{sec:massComparison}

\begin{figure} 
	\includegraphics[width=\columnwidth]{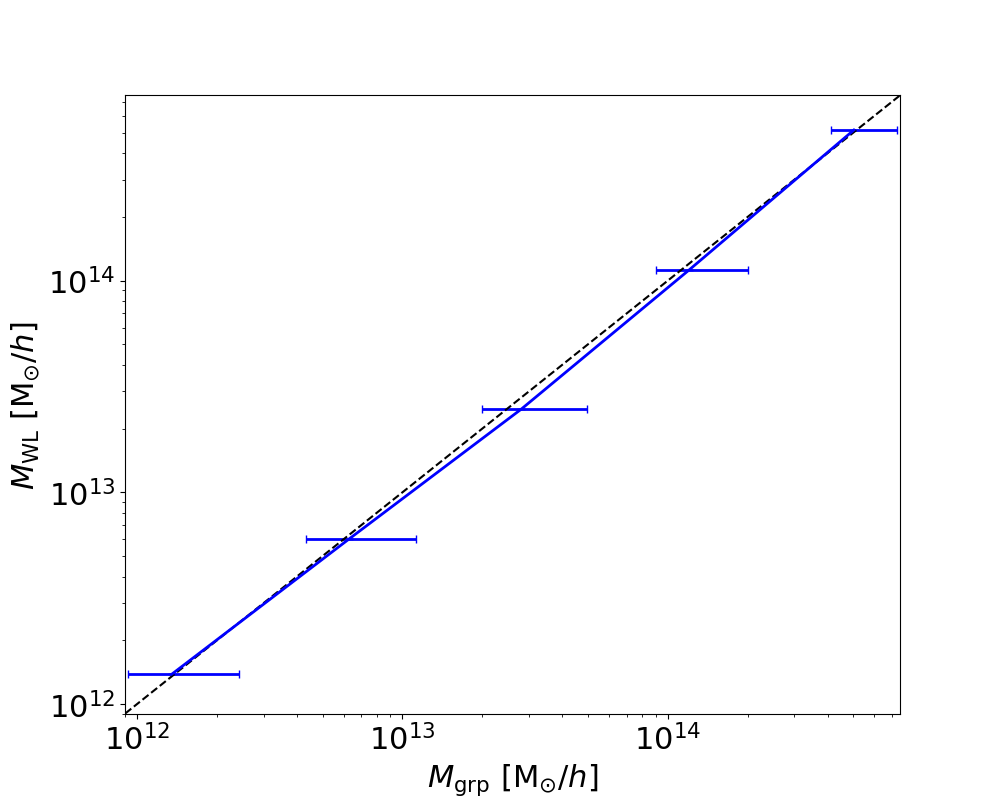}
	\caption{
	Mass comparison between WL fits to those from the group catalog. The group catalog mass has been converted to our adopted cosmology (see Section~\ref{sec:simulation}) and mass definition (virial mass). The horizontal error bars are the $\pm 34$ percentiles from the medians and the vertical error bars are the errors from the WL fits. Note that the WL mass error bars are too small ($\approx 0.02$ dex) to be visible.
	}
	\label{fig:MassComparison}
\end{figure}
We compare the weak lensing masses to the group catalog masses in Figure~\ref{fig:MassComparison}. The weak lensing masses, $M_{\rm WL}$, are obtained by fitting the 2-halo NFW profiles (Equation~\ref{eq:WLMassModel}) to the weak lensing measurements (Figure~\ref{fig:excessSurfaceDensity_fit}), while the halo masses, $M_{\rm grp}$, are from the group catalog used in this work (see Section~\ref{sec:lensCatalog}). We find that the mean ratio between weak lensing versus group catalog virial masses is roughly $0.96$. This is consistent with the results of \citet{Han2015} who found their luminosity based group mass calibration is higher than WL mass in the GAMA catalog.

To make this comparison we first convert the halo catalog masses, $M_{\rm grp}$, from the cosmology used in~\cite{Yang2021} to the one used in this work. This is done by mapping the masses through the peak height. The peak height is used because the early Universe is well described by a random Gaussian field, and the peak height quantifies the mass of a collapsed halo from the fluctuations in the linear density field for a given cosmology. For example, in~\cite{Angulo2010} the peak height is used to convert mass functions between cosmologies. After mapping the masses to the cosmology used in this work, we convert them to $M_{\rm vir}$ by assuming a NFW mass profile with the mass-concentration relation from~\cite{Diemer2019}, because haloes are well described by the NFW profile on these scales.

\section{Lensing bias profile modelling code}
\label{sec:WLCode}
We create a module, inspired by the \textsc{Colossus} package~\citep{Diemer2018}, that predicts the weak lensing measurements based on a given model. Here we follow the procedure in Section~\ref{sec:biasAndWeakLensing} based on the the bias profile (Equation~\ref{eq:bFit}), assuming that the stacked halo densities are spherical.  
We extensively test the module against the CosmicGrowth Simulations with great success and have high confidence in the results. However, we wish to add this section to address a few points about the model itself, and how these are dealt with in our code. 

It is important to note that the bias profile is a form-fitting one and that the measured radii in this work are not parameterized in the model, but derived from it. When the bias parameter priors are wide we sometimes obtain bi-modal posterior distributions that makes the optimal bias parameters difficult to automate. Thus for the results shown in this work we condition the parameter priors based on sub-cluster ($\log M_{\rm vir} < 14 ~{\rm M_{\odot}}/h$) and cluster ($\log M_{\rm vir} \geq 14 ~{\rm M_{\odot}}/h$) masses, where this scale divides bias profiles with troughs from those without (see Figure~\ref{fig:excessSurfaceDensity_fit}. The mass-conditioned prior produces clearer MCMC posteriors for all bias parameters where an optimal median bias profile can be easily determined. 

The bias profile asymptotes at the very central region, $r < 0.06 {\rm Mpc}/h$. This can cause issues when choosing a lower limit of integration for the lensing profiles (see Section~\ref{sec:biasAndWeakLensing}), which we will call $R_{\rm min}'$ in place of $0$.
To allow us to integrate to ever smaller scales, we interpolate the $\Sigma$ profiles (univariate spline) from our model down to smaller scales and compare them with the simulation $\Delta \Sigma(R)$ profiles. From these comparisons we choose an appropriate $R_{\rm min}'$. Regardless of how this might bias the very central region of halo profiles, for example cuspy/core halo profiles, this is not expected to impact the halo profile beyond the central region.

In the module $\Sigma(R)$ is sensitive to the upper integration limit in the form of adding mass-sheets. We will call this upper limit $R_{\rm max}'$, in place of $\infty$ in Section~\ref{sec:biasAndWeakLensing}. The mass-sheet effect is mostly cancelled out in $\Delta \Sigma(R)$, the observable in this work, though there will be a build up of mass in the profile around $\sim R_{\rm max}'$. In our simulations we cut out cubes centered on the haloes in question, to obtain the halo profiles. In this case we find that $R_{\rm max}'$ is optimal, for both the surface and excess surface density profiles, when the maximum integration value is half the box length.
In observations we avoid the mass build-up by choosing $R_{\rm max}'$ as a sufficiently high value, well beyond the observation limit. Though ideally this should be balanced by the number of radial bins and integration time.

\section{Comparing maximum velocity calculation methods}
\label{sec:VmaxDiscussion}
The ratio of the maximum-to-virial circular velocities, $V_{\rm max}/V_{\rm vir}$, is positively correlated with halo concentration~\citep{Gao2007,Angulo2008,Sunayama2016}, where $V_{\rm max}$ is the maximum circular velocity and $V_{\rm vir}$ is the circular velocity at the virial radius.
However, there are differing methods in calculating $V_{\rm max}$. 
In this section we will compare the two different methods in calculating $V_{\rm max}$, which was briefly introduced in the main.
There, we calculate $V_{\rm max}$ using only particles bound to the main \textit{central} subhalo, excluding satellite and unbound particles. In this section we will include ``-cen'' to the labels using this method. An alternative way of calculating $V_{\rm max}$ is to use \textit{all} particles around the halo center, for which we will include ``-all'' to the labels. Note that we use all particles when calculating $M_{\rm vir}$, and thus $V_{\rm vir}$, and all calculations are done at $z=0.253$.

\begin{figure} 
	\includegraphics[width=\columnwidth]{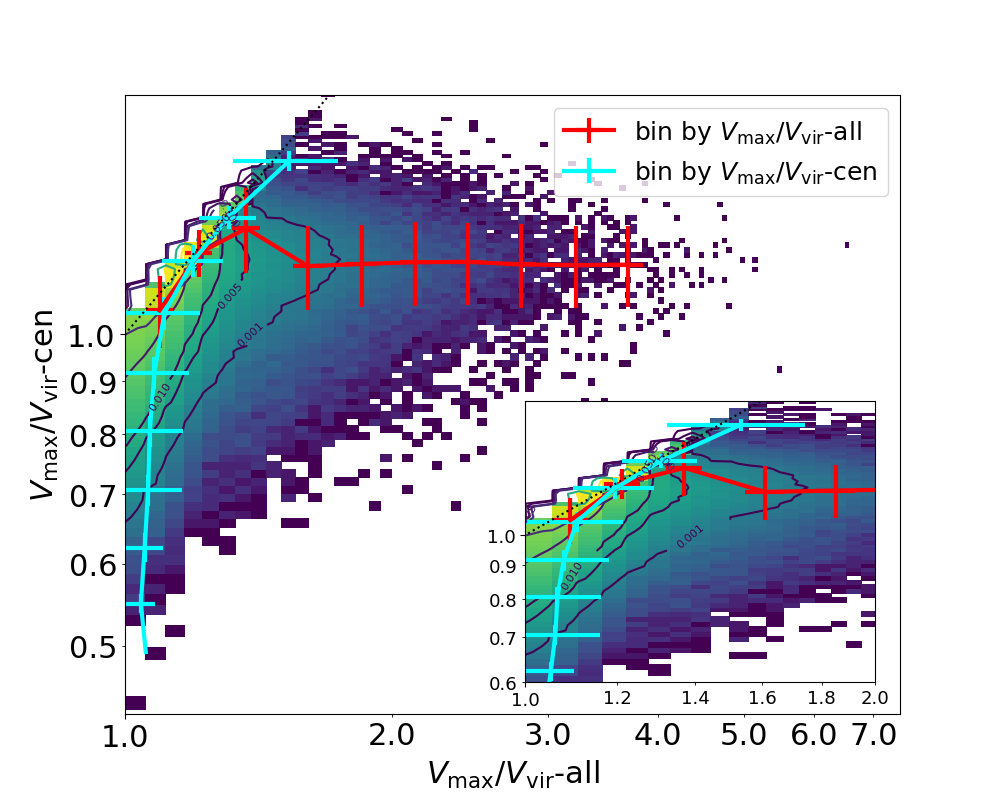}
	\caption{
	Comparison between the $V_{\rm max}/V_{\rm vir}$ calculation methods. The colored contour represents the normalized density of haloes, while the labelled line contours are isodensities. The line and errorbars represent the mean and standard deviation within each bin, respectively, while the red and cyan colors represent the distribution when binned by $V_{\rm max}/V_{\rm vir}$-all and $V_{\rm max}/V_{\rm vir}$-cen methods, respectively. The bottom right subplot is a zoom-in of the full image. For reference we include $y=x$ as the black dotted line in both panels. 
	}
	\label{fig:VmaxComparison}
\end{figure}
In Figure~\ref{fig:VmaxComparison} we compare the $V_{\rm max}$ calculation methods, in the form of $V_{\rm max}/V_{\rm vir}$, one of the simulation proxies for concentration that is expected to be related to the mass accretion history of a halo~\citep{Wechsler2002,Zhao2003b,Lu2006,Zhao2009,Ludlow13,Wang2020}.
The lines represent the distribution when binned, into ten logarithmic bins, by $V_{\rm max}/V_{\rm vir}$-all (red) or $V_{\rm max}/V_{\rm vir}$-cen (cyan).
This shows that the maximum circular velocity calculation methods are not equivalent. In the main text we take the upper and lower quartiles of the concentration proxy, $V_{\rm max}/V_{\rm vir}$, and so the quartiles will sample completely different set of haloes depending on the method that $V_{\rm max}$ is calculated.

\begin{figure*}
     \centering
     \begin{subfigure}[b]{\columnwidth}
         \centering
         \includegraphics[width=\columnwidth]{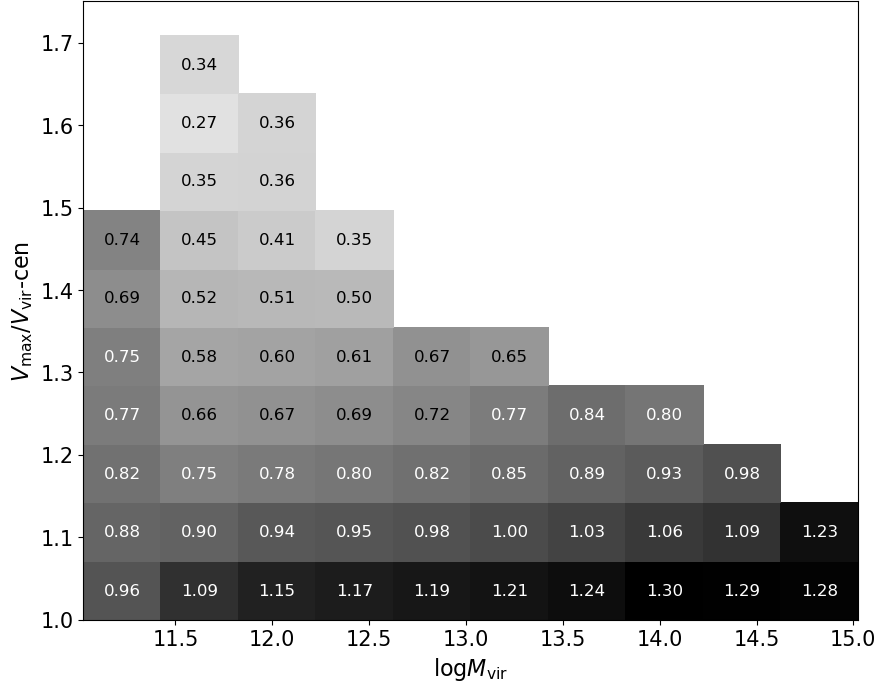}
         \label{fig:mar_logMvirVsVmax_Vvir_hostVsSubX1X2_zoom}
     \end{subfigure}
     \hfill
     \begin{subfigure}[b]{\columnwidth}
         \centering
         \includegraphics[width=\columnwidth]{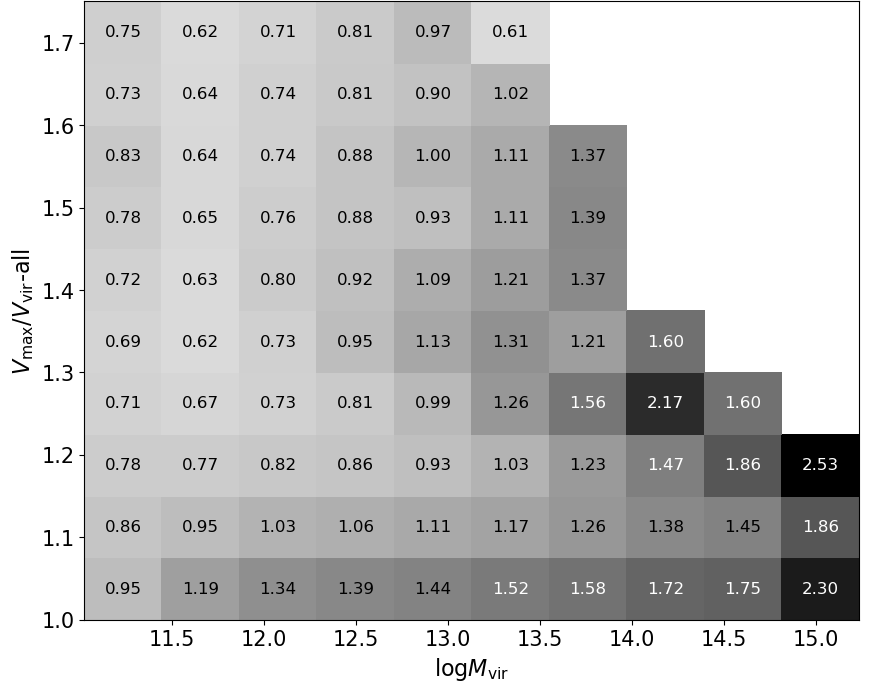}
         \label{fig:mar_logMvirVsVmax_Vvir_subVsHostX1X2_zoom}
     \end{subfigure}
        \caption{
        The zoomed-in pixelated mass accretion rate values when haloes are binned by halo mass and $V_{\rm max}/V_{\rm vir}$-cen (left) and $V_{\rm max}/V_{\rm vir}$-all (right). 
        }
        \label{fig:mar_Vmax_comparison} 
\end{figure*}

In Figure~\ref{fig:mar_Vmax_comparison} we plot the mass accretion rate (MAR) in each bin when haloes are binned by halo mass and $V_{\rm max}/V_{\rm vir}$ for $V_{\rm max}/V_{\rm vir}$-cen and $V_{\rm max}/V_{\rm vir}$-all calculation methods on the left and right panels, respectively. The MAR is obtained by using the mass evolution history of each halo, using \textsc{HBT+}~\citep{Han2012,Han2018}. From the history catalog we use the halo mass and corresponding redshifts at $10 \%$ mass intervals of the final mass. We calculate the MAR for each halo by $\frac{\log M_{\rm bound}(z) - \log M_{\rm bound}(z') }{\log a(z) - \log a(z')}$, where $M_{\rm bound}$ represents the bound mass of a halo, and $z$ and $z'$ are the redshifts bracketing the redshift of interest (in this case at $z=0.253$). Note that the $V_{\rm max}/V_{\rm vir}$-all distribution is much wider (see Figure~\ref{fig:VmaxComparison}), so we zoom in on overlapping regions.

For a given mass, the $V_{\rm max}/V_{\rm vir}$-cen method has a relatively clearer trend with MAR than the $V_{\rm max}/V_{\rm vir}$-all method counterpart. Thus, we believe that binning our simulated haloes by the concentration proxy $V_{\rm max}/V_{\rm vir}$-cen has a stronger relationship with the MAR.
It is for this reason that we use only the central subhalo particles when calculating $V_{\rm max}$ in the main text.

\bsp	
\label{lastpage}
\end{document}